\documentclass[twocolumn]{aastex631}

\newcommand{\irex}{Institut Trottier de Recherche sur les Exoplanètes, Département de Physique, Université de Montréal, 1375 Avenue Thérèse-Lavoie-Roux, Montreal, QC H2V 0B3, Canada}
\newcommand{\planet}{HIP~65Ab}

\graphicspath{{./}{figures/}}

\begin{document}

\title{A sub-solar metallicity on the ultra-short period planet HIP~65Ab}

\correspondingauthor{Luc Bazinet}
\email{luc.bazinet@umontreal.ca}

\author[0000-0003-3181-5264]{Luc Bazinet}
\affiliation{\irex}

\author[0000-0002-8573-805X]{Stefan Pelletier}
\affiliation{\irex}
\affiliation{Observatoire astronomique de l'Université de Genève, 51 chemin Pegasi 1290 Versoix, Switzerland}

\author[0000-0001-5578-1498]{Bj\"{o}rn Benneke}
\affiliation{\irex}

\author[0000-0002-1206-1930]{Ricardo Salinas}
\affiliation{Departamento de F\'isica y Astronom\'ia, Universidad de La Serena, Av. Juan Cisternas 1200 Norte, La Serena, Chile}

\author[0000-0001-7875-6391]{Gregory N. Mace}
\affiliation{Department of Astronomy and McDonald Observatory, The University of Texas at Austin, 2515 Speedway, Stop C1400, Austin, TX 78712, USA}
    


\begin{abstract}

Studying and understanding the physical and chemical processes that govern hot Jupiters gives us insights on the formation of these giant planets.
Having a constraint on the molecular composition of their atmosphere can help us pinpoint their evolution timeline. Namely, the metal enrichment and carbon-to-oxygen ratio can give us information about where in the protoplanetary disk a giant planet may have accreted its envelope, and subsequently, indicate if it went through migration.
Here we present the first analysis of the atmosphere of the hot Jupiter \planet. Using near-infrared high-resolution observations from the IGRINS spectrograph,
we detect H$_2$O and CO absorption in the dayside atmosphere of \planet.
Using a high-resolution retrieval framework, we find a CO abundance of log(CO) = $-3.85^{+0.33}_{-0.36}$, which is slightly under abundant with expectation from solar composition models. We also recover a low water abundance of log(H$_{2}$O) = $-4.42\pm{0.18}$, depleted by 1 order of magnitude relative to a solar-like composition.
Upper limits on the abundance of all other relevant major carbon- and oxygen-bearing molecules are also obtained.
Overall, our results are consistent with a sub-stellar metallicity but slightly elevated C/O.
Such a composition may indicate that \planet\ accreted its envelope from beyond the water snowline and underwent a disk-free migration to its current location.
Alternatively, some of the oxygen on \planet\ could be condensed out of the atmosphere, in which case the observed gas-phase abundances would not reflect the true bulk envelope composition.

\end{abstract}



\section{Introduction} \label{sec:intro}

Hot Jupiters offer the best opportunity to study atmospheric properties of exoplanets. With their large radius and elevated temperatures, they are ideal for emission spectroscopy. With no analogue in our Solar System, these extreme worlds are interesting to study from a planet formation perspective.
Specifically, measuring elemental abundance ratios in hot Jupiter atmospheres can shed insight into their evolutionary and migratory pasts \citep{oberg_effects_2011, madhusudhan_toward_2014, lothringer_new_2021}.
These key elements include oxygen and carbon, the two most important metal building blocks of planets \citep{line_solar_2021}.
Water and other oxygen-bearing molecules in Jupiter and Saturn's atmosphere are condensed due to their cold temperature. Finding the elemental abundances in hot gaseous planets is considerably easier, as their higher temperatures enables water to remain in gas-form.

High-resolution cross-correlation spectroscopy \citep[HRCCS,][]{snellen_orbital_2010, brogi_signature_2012, birkby_spectroscopic_2018} coupled with a Bayesian retrieval framework has proven to be a powerful tool for the characterization of hot Jupiters \citep[e.g.,][]{brogi_retrieving_2019, gandhi_hydra-h_2019, gibson_detection_2020, gibson_relative_2022}. HRCCS uses data from high resolution ($R \gtrsim 25,000$) spectrographs to distinguish between planetary lines and the undesirable telluric and stellar lines.
These different spectral lines can be untangled because the planetary lines' radial velocity undergo several km\,s$^{-1}$ shifts during continuous hours-long observations, while the telluric and stellar lines remain stationary or quasi-stationary. 
The dayside temperature structure and volume mixing ratios of atmospheric constituents dictate the observed line shapes and contrasts. From observing the emission spectrum of a planet, we can therefore infer the underlying physical properties of its atmosphere using a Bayesian retrieval framework coupled with an atmospheric forward model \citep[e.g.,][]{line_solar_2021}.

Just recently discovered in 2020, \planet\ is a massive ($M_{p} = 3.213 \pm{0.078}$\,$M_{J}$) hot Jupiter ($T_{\mathrm{eq}} = 1411\pm{15}$\,K) on an ultra short 0.98 day orbit around its K4V ($K_{mag} = 8.29$) host star \citep{nielsen_three_2020, wong_systematic_2020}. Being a grazing transiting planet with an impact parameter $b = 1.17_{-0.08}^{+0.10}$, its radius is relatively poorly constrained ($R_{p} = 2.03_{-0.49}^{+0.61}$\,$R_{J}$).
With its short orbital period, relatively large radius and cold host star, it is a prime target for atmospheric characterization using high-resolution spectroscopy.

The atmosphere of \planet\ has yet to be studied in great details. Here, we use HRCCS with a Markov Chain Monte Carlo based retrieval framework to find the abundances of major carbon- and oxygen-bearing molecules. This work offers a first look into the dayside atmosphere of \planet.

In Section \ref{sec:obs}, we discuss the observations of \planet\ with IGRINS. Section \ref{sec:datareduc} presents the data reduction steps and the telluric lines removal process. The modelling of \planet's atmosphere and the retrieval setup are discussed in Section \ref{sec:model}. 
In Section \ref{sec:res}, we present the results from the cross-correlations and retrievals.
These results are discussed in Section \ref{sec:discussion}.
We conclude this work in Section \ref{sec:conclusion}.

\section{Observations} \label{sec:obs}

\begin{figure}
    \centering
    \includegraphics[width=\linewidth]{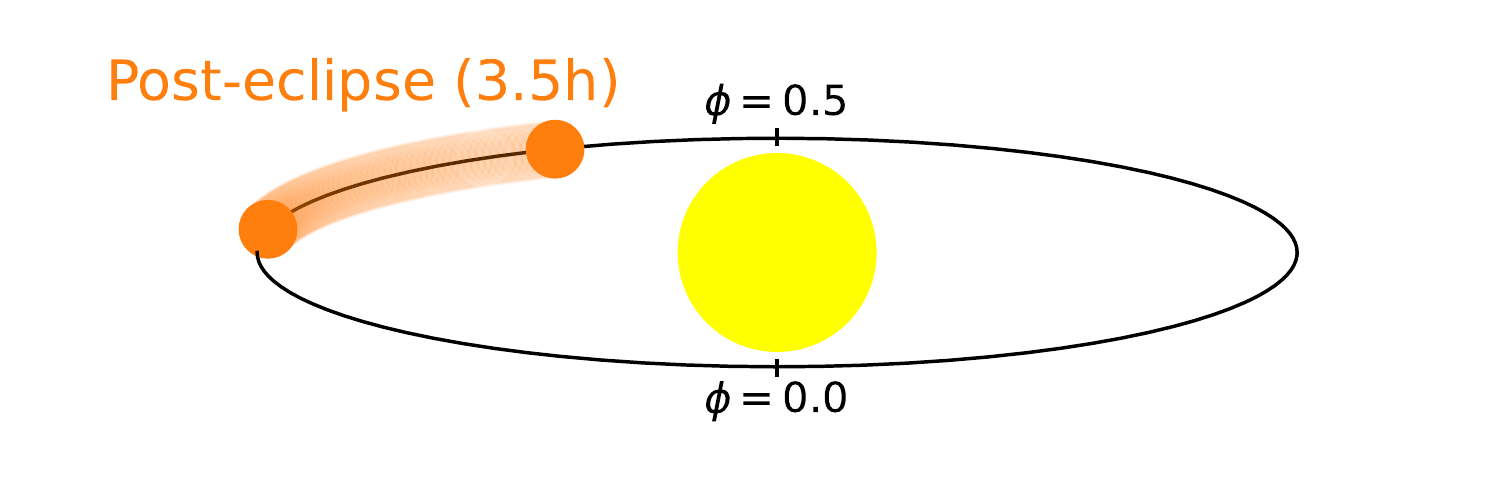}
    \caption{To scale diagram of the phase cover of our \planet\ observations. The phase covered is from $\phi = 0.57$ to $0.717$ where $\phi = 0$ is the center of the transit and $\phi = 0.5$ is the middle of the secondary eclipse. The 3.5 hours observation used in this analysis is of the post-eclipse (western) dayside of \planet.}
    \label{fig:obs}
\end{figure}

We observed \planet\ with the Immersion GRating INfrared Spectrometer (IGRINS) \citep{park_design_2014, mace_igrins_2018} for 3.5 hours during the night of 7 July 2022 (Program ID GS-2022A-Q-137, PI: Pelletier). Our data set composed of 56 exposures of the western dayside hemisphere of the planet with orbital phases from $\phi = 0.57$ to $0.717$ (Figure \ref{fig:obs}). Exposures were taken in an ABBA nodding pattern to reduce background contamination. The exposure time of each AB pair was 140 seconds, with an average SNR per exposure of about 63.

IGRINS is a high resolution spectrograph ($R = 45,000$) installed on the Gemini South 8.1\,m telescope in Chile. It observes in the near infrared in two continuous bands between 1.43 and 2.52\,µm, corresponding to the photometric $H$ and $K$ bands, respectively. This wavelength range contains strong molecular bands from major carbon- and oxygen-bearing molecules, such as H$_{2}$O, CO, CH$_{4}$ and CO$_{2}$ \citep{gandhi_molecular_2020}.

\section{Data reduction and telluric removal} \label{sec:datareduc}

The raw data was processed with the IGRINS Pipeline Package \citep{mace_igrins_2018, lee_plp_2016} which extracts the 1D spectra and does an initial wavelength calibration. To proceed with our analysis, we bundled the IGRINS pipeline products in a data cube of 56 (exposures) $\times$ 54 (spectral orders) $\times$ 2048 (pixels).

We noticed a $\sim$1 km\,s$^{-1}$ drift of the wavelength solution throughout the continuous observing sequence, which has been reported in previous works using IGRINS \citep[e.g.,][]{brogi_roasting_2023, line_solar_2021}.
To fix this issue, we used a similar technique as in \citet{line_solar_2021}, however, instead of correcting each exposure individually, we corrected the drift using a linear radial-velocity shift in time.
More precisely, we used the same method as in \citet{line_solar_2021} to find the shift of each spectral order for all exposures. We then found the best fitting line of the shift as a function of exposure. Finally, we shifted the whole time series according to this linear fit.
Using a linear fit as opposed to shifting each exposure individually serves to prevent any sporadic shifts to be added to the data in certain outlier cases.
This correction was done separately for spectral orders in the $H$ and $K$ bands, where we found a shift of 0.3 and 1.3 km\,s$^{-1}$ throughout the night, respectively.


Following the initial processing of the raw data, we want to remove the contribution from \planet's host star and the Earth's atmosphere to only keep the planet spectrum buried in some intrinsic noise. To do this, we closely followed the data detrending procedure described in \citet{pelletier_where_2021, pelletier_vanadium_2023}.

As a first step, we corrected the pixels that deviate significantly from their spectral channel. Pixels that are more than 5\,$\sigma$ away from the temporal mean are flagged as bad pixels. If these bad pixels form a group of 3 or less pixels, they are interpolated using neighbouring pixels. Bigger groups of adjacent bad pixels are completely masked.

Afterwards, we masked the wavelengths regions with high telluric contamination. More specifically, the masking was applied to the following regions: 1.43 to 1.4465\,µm, 1.7995 to 2.03\,µm and 2.48 to 2.52\,µm, corresponding to the edges of the $H$ and $K$ bands. Another mask was introduced to remove the deep features that dip below 70\% of the continuum flux. We also masked 200 pixels on both edges of each order, where the blaze transmittance is lower, to remove edge residual effects observed in some cases.
Finally, we discard any remaining spectral orders where more than 80\% of the data points are masked by the aforementioned steps. This resulted in the removal of 10 of the 54 spectral orders, all of which are at the edges of the $H$ and $K$ bands where telluric absorption is strongest.

\begin{figure}[t]
    \centering
    \includegraphics[width=\linewidth]{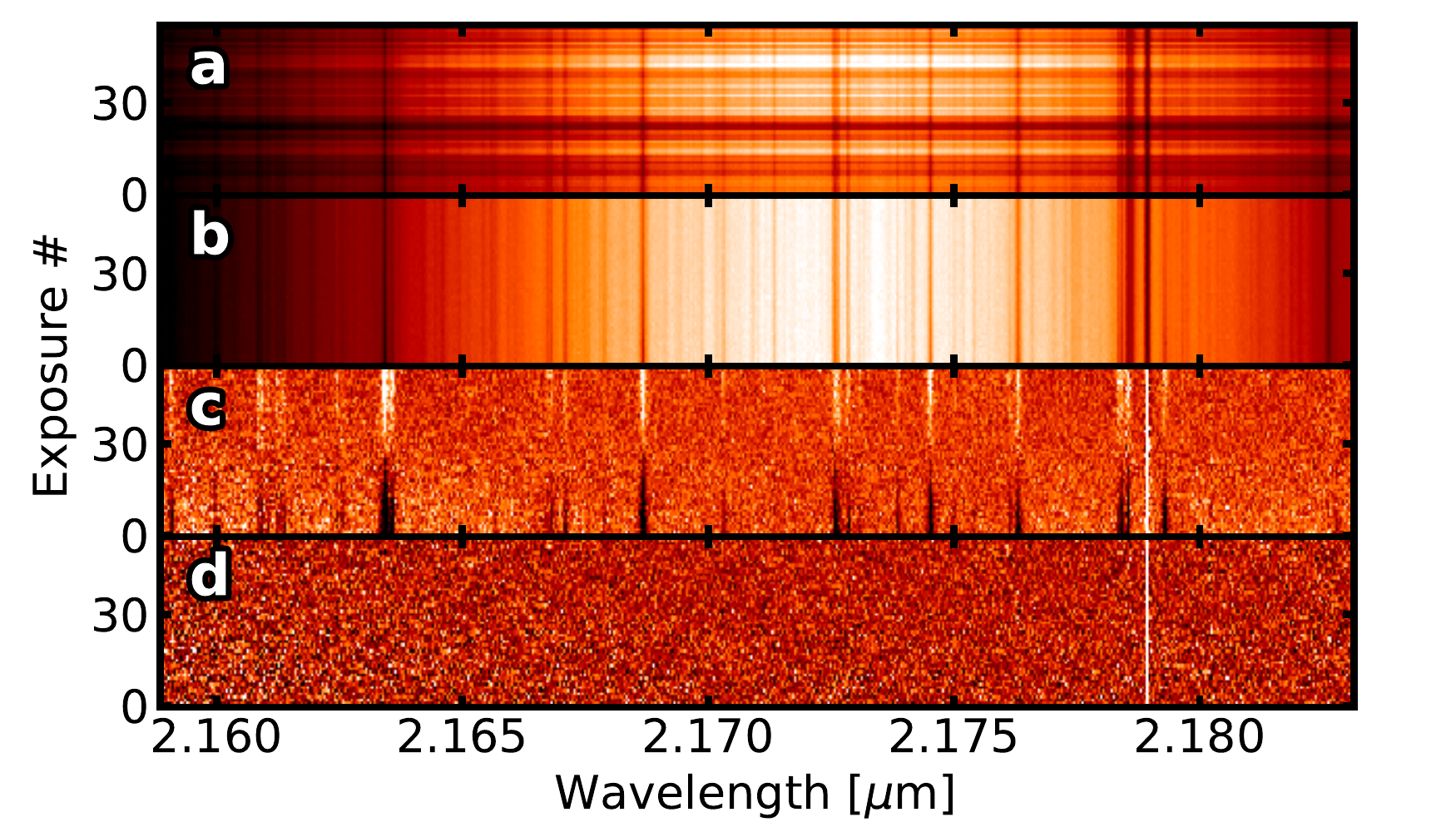}
    \caption{Overview of the data detrending steps applied to the observed 3.5 hour IGRINS spectroscopic time series targeting \planet's dayside atmosphere. \textbf{a)} The extracted data from the IGRINS pipeline of an example spectral order in the $K$ band.
    \textbf{b)} The same data, but continuum alignment. \textbf{c)} After the fitted median spectra is removed from each exposure. \textbf{d)} With three principal components removed and noisy spectral channels masked. This last panel is fed to the cross-correlation and retrieval frameworks. The applied steps help remove undesirable telluric and stellar contributions, leaving behind the underlying planetary signal buried in noise.}
    \label{fig:reductionsteps}
\end{figure}

Because of the varying observing conditions between each exposure (e.g., due to airmass, seeing, throughout variations), the photon count for each pixel varies greatly throughout the time series (Figure~\ref{fig:reductionsteps}, panel a). 
To bring each observed spectrum to the same continuum level, we used the same procedure as in \citet{pelletier_where_2021}. However, we used a 51 pixel width box filter modulated with a Gaussian filter of 100 pixel standard deviation \citep{gibson_revisiting_2019, gibson_detection_2020}.
This process corrects for any continuum variations between exposures and enables a differential analysis to be performed on the time series of spectra (Figure~\ref{fig:reductionsteps}, panel b).
A bigger filter has been also tested as this has been found to sometimes influence retrieval results \citep{gibson_relative_2022}. However, in our case, we retrieve similar results in both cases.

At this stage, remaining telluric lines are still the dominating variance contained in the spectra. However, these are stationary in the earth's frame of reference. Another major source of variance in the data is the spectral lines produced by the host star HIP~65A. These stellar lines are shifted by less than a pixel throughout the 3.5 hour observation caused mainly by the barycentric movement of the earth and reflex motion of \planet. For comparison, \planet's shift is about 55 pixels ($\sim$ 110\,km\,s$^{-1}$) throughout the 3.5\,h observation due to its short orbital period.

The stability of telluric and stellar lines relative to the rapidly moving spectral lines emitted from \planet's atmosphere can be used to remove these unwanted contributions while leaving the planetary signal mostly unaffected. For this, we first divide the spectra by a second order fit of the median spectrum of each order. This is done individually for each exposure.
Without change in airmass, stellar activity and other time dependant changes, this reduction would theoretically be enough to remove the telluric and stellar line while keeping the planetary signal intact. Unfortunately, such variations and noises are present, leaving residuals that still overwhelm the planetary signal (Figure~\ref{fig:reductionsteps}, panel c). 


\begin{figure}
    \centering
    \includegraphics[width=\linewidth]{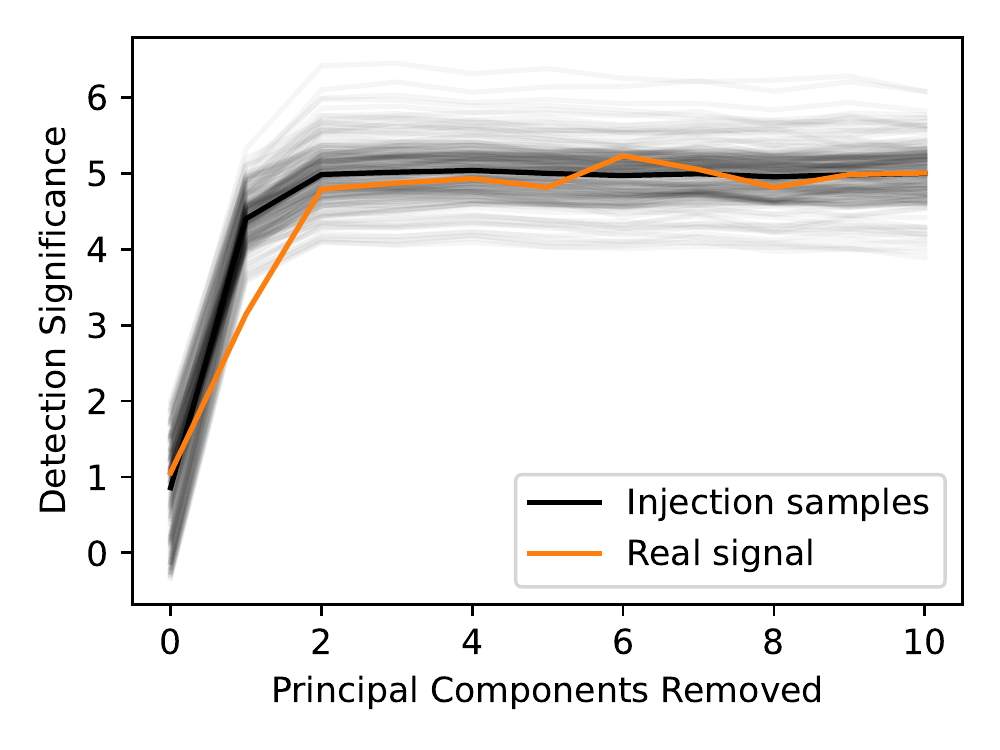}
    \caption{Evolution of the cross-correlation detection strength as a function of number of principal components removed for both the real signal, and 200 injected samples.
    The orange line represents the detection of the real signal from \planet\ observed in the data 
    for different number of removed principal components.
    Each grey line represents an injection-recovery test (black line = the average), where we inject an atmospheric model at random orbital parameters away from the expected signal in $K_{p}-V_{\mathrm{sys}}$ space and measure the strength at those parameters as a function of components removed. 
    The strength of the injected model was calibrated to roughly match the real signal line. 
    Both the real signal and the injected signal exhibit similar behavior when varying the number of principal components removed.  In all cases, at least 2 principal components need to be removed to well-reveal the planetary signal, which then remains persistent up to at least 10 components removed.
    In our final analysis, we opted to remove 3 components.}
    \label{fig:PCAvsdetection}
\end{figure}

Next, principal component analysis (PCA) is used to remove the remaining contamination from telluric and stellar residuals. PCA finds the orthogonal components of a data set with the most variance. The first few components of the PCA will be orientated toward the telluric and stellar lines as they are the dominating source of time dependant variance. 
The planetary lines shift in wavelength by a considerable amount compared to telluric and stellar lines, therefore, the PCA will not prioritise these.
However, removing an excessive amount of components can eventually chip away \planet's signal. Finding the optimal number of components removed is therefore a balance between the complete removal of the undesirable lines of telluric or stellar nature, and the preservation of lines from \planet's atmosphere \citep{birkby_discovery_2017, cabot_robustness_2019, cheverall_robustness_2023, smith_combined_2024}.
As done with other high-resolution cross-correlation analyses \citep[e.g.,][]{holmberg_first_2022}, we calculated the detection strength as a function of number of principal components removed for the real signal and 200 injected signals at random locations in $K_p$-$V_{\mathrm{sys}}$ space away from the real signal (Figure \ref{fig:PCAvsdetection}). We found that the signal is best recovered if at least 2 principal components are removed, and changes little if up to 10 are removed. For our analysis, we opted to remove 3 components.

Finally, a mask is applied to any remaining noisy spectral channels that deviate more than 5\,$\sigma$ of their spectral order. This final step removed about 0.3\% of the data.

The results of all these detrending steps is a time series of \planet's spectra dominated by noise (Figure~\ref{fig:reductionsteps}, panel d). These spectra are fed to the cross-correlation and retrieval frameworks to, ultimately, characterise \planet's atmosphere. 

\section{Modelling and retrieval setup} \label{sec:model}

In order to detect chemical species, the data must be cross-correlated with appropriate atmospheric models. Here, we use the SCARLET framework \citep{benneke_atmospheric_2012, benneke_how_2013, benneke_strict_2015, knutson_featureless_2014, kreidberg_clouds_2014, benneke_water_2019, benneke_sub-neptune_2019, pelletier_where_2021, pelletier_vanadium_2023} to generate models of \planet's dayside thermal emission. SCARLET generates line-by-line high-resolution ($R = 250,000$) emission spectra given the temperature-pressure profile, the composition and the cloud structure of a planet.
Molecular opacities used for this analysis are H$_{2}$O \citep{polyansky_exomol_2018, barber_high-accuracy_2006}, CO \citep{rothman_hitemp_2010, li_rovibrational_2015}, CH$_4$ \citep{rothman_hitemp_2010, hargreaves_accurate_2020}, CO$_2$ \citep{yurchenko_exomol_2020}, SiO \citep{yurchenko_exomol_2022}, HCN \citep{harris_improved_2006, barber_exomol_2014} and C$_2$H$_2$ \citep{chubb_exomol_2020}.
Collision-induced absorption by H$_{2}$-H$_{2}$ and H$_{2}$-He \citep{borysow_collision-induced_2002} are also considered.

For each generated model, three sources of broadening are then taken into account: rotational, exposure, and instrumental. 
Rotational broadening is due to the line-of-sight spin of the planet, $V_{\mathrm{rot}} = v\mathrm{sin}(i)$ (see Eq.\ 4 from \citep{reiners_feasibility_2002}, assuming $\epsilon = 1$), causing different parts of the \planet's dayside rotating towards (blueshifted) and away (redshifted) from the observer. 
Exposure broadening considers the broadening effect caused by the radial velocity change of the planet during a given exposure.
\planet\ has an ultra-short period of 0.98 days, therefore it travels a non-negligible distance during an exposure. This creates a detectable blurring of the molecular lines in the data that needs to be considered in the modelling. The average blurring per AB pair by this source is about 1.2 km\,s$^{-1}$. 
Finally, to account for the instrumental broadening, the spectrum is convolved with a Gaussian profile matching IGRINS' spectral resolution ($R = 45,000$).

\begin{figure}
    \centering
    \includegraphics[width=\linewidth]{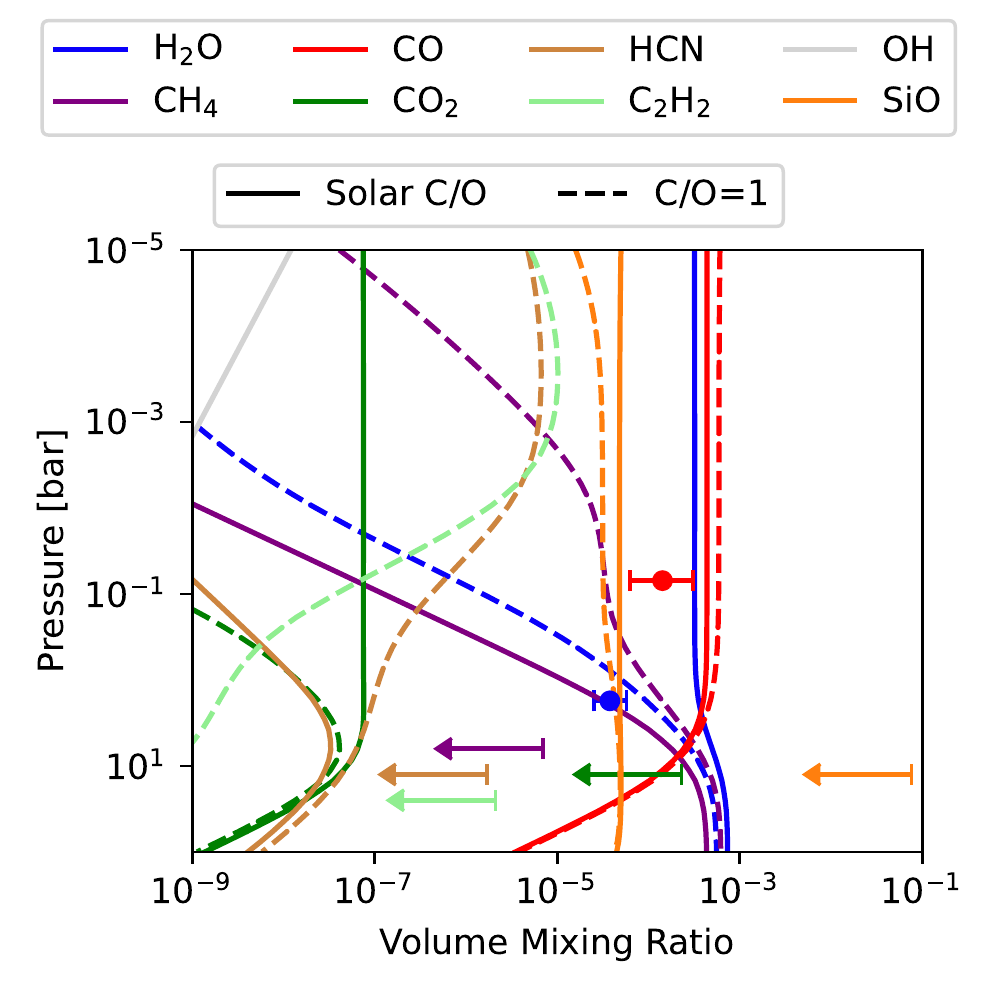}
    \caption{Mixing ratios of notable carbon- and oxygen-bearing molecules in chemical equilibrium, as predicted by \texttt{FastChem 2} \citep{stock_fastchem_2022, stock_fastchem_2018}, compared to retrieved abundances (Section \ref{sec:retrievalres}). The chemical equilibrium models are calculated for a solar metallicity with a uniform temperature profile of 1411\,K, the equilibrium temperature of \planet. The solid lines are the mixing ratios for a model with solar C/O, while the dashed lines represent a C/O of 1.
    The retrieved H$_2$O and CO abundances are illustrated with their 1$\sigma$ uncertainties. The other abundance constraints are shown as their 3$\sigma$ upper limits.
    All shown molecules are included in our analysis, with the exception of OH, which is never expected to have measurable abundances at the pressures probed.
    The retrieved CO is slightly under-abundant compared to equilibrium models. The H$_2$O is also depleted relative to these solar abundance model predictions.
    SiO is poorly constrained due to it having almost no opacity in the IGRINS bandpass.
    Overall, our results are consistent with \planet\ having a sub-solar metallicity, and show a preference towards a super-solar C/O.
    }
    \label{fig:mix_ratio}
\end{figure}

As a first approximation of the composition of \planet, we used SCARLET to produce chemical equilibrium models of \planet\ assuming a self-consistent temperature structure with a zero-albedo equilibrium temperature of 1411\,K. At each atmospheric layer, SCARLET uses the thermochemical equilibrium code \texttt{FastChem 2} \citep{stock_fastchem_2022, stock_fastchem_2018} to compute the volume mixing ratios (VMR) of chemical species given the temperature, the pressure and the elemental abundances of a volume of gas. We produced two models: one with a solar metallicity and C/O, and the other at a solar metallicity but with a C/O of 1 (Figure \ref{fig:mix_ratio}). In both cases, the models predicts that the molecular composition is dominated by H$_2$O, CO, CH$_4$ and SiO in the low atmosphere. As \planet's temperature is significantly less than the dissociation temperature of water \citep{parmentier_thermal_2018}, OH is never predicted to be in measurable abundances (Figure \ref{fig:mix_ratio}). In our search for molecular species in \planet's atmosphere, we therefore looked for signals from H$_2$O, CO, CH$_4$, SiO as well as CO$_2$, HCN and C$_2$H$_2$.

\subsection{Retrieval setup} \label{sec:retrieval}

To be able to retrieve the atmospheric parameters on \planet, the spectrum needs to be Doppler shifted according to the relative speed of \planet\ at each exposure. The equation that dictates the radial velocity shift $V_{p}(t)$ at a given time for a planet of zero eccentricity, as is the case for \planet, is given as:
\begin{equation} \label{eq:dopplershift}
     V_{p}(t) = K_{p}\sin(2\pi\phi(t)) + V_{\mathrm{sys}} + V_{\mathrm{bary}}(t),
\end{equation}
where $\phi(t)$ is the orbital phase of the planet at time $t$, $V_{\mathrm{bary}}(t)$ is the barycentric radial velocity of the Earth, $K_{p}$ is the Keplerian velocity and $V_{\mathrm{sys}}$ is the systemic velocity.

We use a Bayesian analysis to characterize \planet's atmosphere and constrain its orbital and atmospheric parameters. The Markov Chain Monte Carlo code \texttt{emcee} \citep{foreman-mackey_emcee_2013} was used for the retrieval of atmospheric and orbital parameters.
Volume mixing ratios (VMRs) of H$_{2}$O, CO, CH$_{4}$, C$_{2}$H$_{2}$, CO$_{2}$, HCN and SiO were fitted. 
The orbital parameters $K_{p}$ and $V_{\mathrm{sys}}$ were also fitted. To find the temperature profile of \planet, we fitted 12 points evenly spaced in the log pressure space between 0.01 mbar and 100 bar. The prior used on the temperature points is described in \citet{pelletier_where_2021} (See their Eq.\ 11) where we used a smoothing parameter $\sigma_{s}$ of 350\,K\,dex$^{-2}$.
An optically-thick grey cloud deck is also considered in the retrieval with its cloud-top pressure $P_{c}$ being fitted. The priors on all molecular abundances were log-uniform from $10^{-12}$ to 1, 
with the added condition that mixing ratios queried by the retrieval summing to more than unity are rejected from the prior space \citep{benneke_atmospheric_2012}. H$_2$ and He is then used as a filler gas such that the sum of all mixing ratios is equal to one.
We adopted a log-uniform prior for the cloud-top pressure of the grey cloud between 0.01 mbar and 100 bar.
A uniform prior is used on the parameter $K_{p}$ between $\pm{50}$\,km\,s$^{-1}$ of its literature value \citep[192.7 km\,s$^{-1}$, calculated with planetary parameters given in][]{nielsen_three_2020}. 
$V_{\mathrm{sys}}$'s uniform prior is also between $\pm{50}$\,km\,s$^{-1}$ of its known value \citep[21.04 km\,s$^{-1}$,][]{gaia_collaboration_vizier_2018}.

The rotational broadening was also fitted with its uniform prior being between 5.18 and 20\,km\,s$^{-1}$. The lower bound of 5.18 was chosen to be the equatorial rotational velocity assuming a planetary radius of 1\,R$_J$.
This choice is motivated by the fact that a radius of 1\,R$_J$ or lower for \planet\ is nonphysically low based on its mass of 3.213 M$_J$.
We expect the retrieved broadening to be close to the tidally-locked equatorial rotation of \planet\ which is about 10.5\,km\,s$^{-1}$, assuming a radius of 2.03\,R$_J$.





The likelihood treatment used in this analysis is based on the framework of \citet{gibson_detection_2020}. The cross-correlation function is given by
\begin{equation} \label{eq:CCF}
         \mathrm{CCF}= \sum^{N}_{i=1} \frac{d_{i} m_{i}}{\sigma^{2}_{i}},
\end{equation}
where $m$ is the planetary atmospheric model, $d$ is the cleaned data (Figure~\ref{fig:reductionsteps}, panel d), $\sigma$ is the uncertainty on the data as calculated in \citet{gibson_detection_2020}, and $N$ is the number of data points. The CCF-to-likelihood mapping is then of the form
\begin{equation} \label{eq:likelihood}
         \ln\mathcal{L} = -\frac{N}{2} \ln\left[\frac{1}{N}\left(\sum^{N}_{i=1} \frac{d_{i}^2+\alpha^2 m_{i}^2}{\sigma^{2}_{i}}-2\alpha\mathrm{CCF}\right)\right],
\end{equation}
where 
$\alpha$ is a factor for the uncertainty in the scaling of the model \citep{brogi_retrieving_2019, gibson_detection_2020}. 
As done by previous works and supported by the poorly constrained radius, we opted to fit this parameter. The prior used on $\alpha$ is a log-uniform between 10$^{-2}$ and 10$^2$.

\section{Results} \label{sec:res}

\subsection{Cross-correlation results} \label{sec:ccfres}

To search for the chemical inventory of \planet, we start by cross-correlating molecular models with the detrended data following Eq.\ \ref{eq:CCF}.
The cross-correlation results are phase-folded at each point in a $K_{p}-V_{\mathrm{sys}}$ grid according to Eq.\ \ref{eq:dopplershift}. 

\begin{figure}
    \centering
    \includegraphics[width=\linewidth]{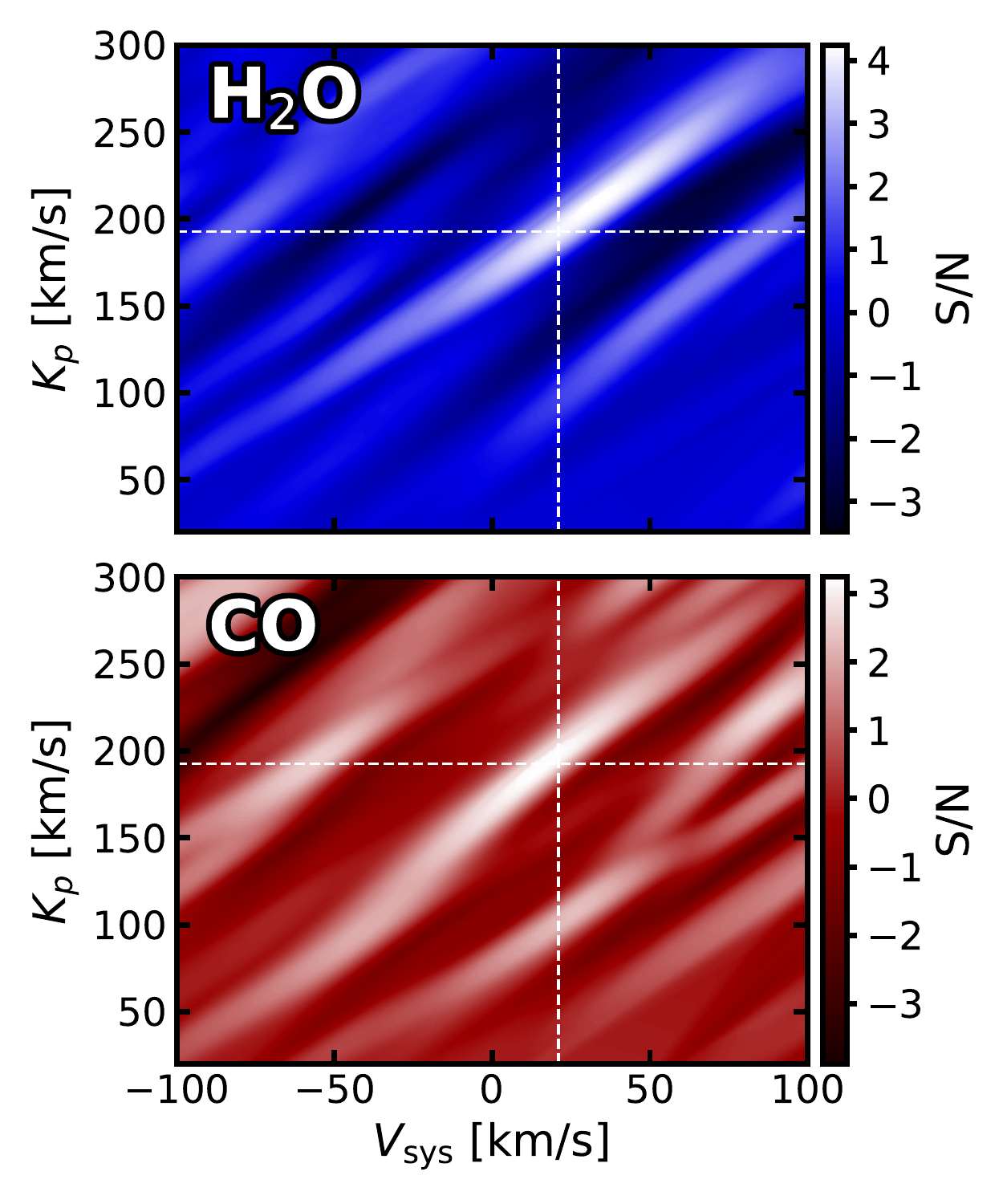}
    \caption{
    $K_{p}-V_{\mathrm{sys}}$ cross-correlation signal-to-noise maps for the observed water (top) and carbon monoxide (bottom) signals in \planet's dayside atmosphere. The white dotted lines indicate the known orbital parameters \citep[$K_{p} = 192.7$\,km\,s$^{-1}$ and $V_{\mathrm{sys}} = 21.04$\,km\,s$^{-1}$,][]{nielsen_three_2020, gaia_collaboration_vizier_2018}. The maximum cross-correlation peaks (4.2 and 3.2 S/N for H$_{2}$O and CO, respectively) occur close to the expected $K_{p}$ and $V_{\mathrm{sys}}$.}
    \label{fig:detect_ccfs}
\end{figure}

We detect H$_{2}$O and CO absorption features at signal-to-noise ratios of 4.2 and 3.2, respectively, with maxima in $K_{p}-V_{\mathrm{sys}}$ space near the expected orbital parameters (Figure \ref{fig:detect_ccfs}). 
In comparison to similar IGRINS observations of the dayside of hot Jupiter WASP-77Ab \citep{line_solar_2021}, the signal strengths are relatively weak considering that \planet\ has the highest Eclipse Spectroscopy Metric \citep[ESM,][]{kempton_framework_2018} at 2.2\,µm to date, if assuming a radius of 2.03 $R_{J}$.
This suggests that \planet\ likely has a radius on the lower estimate of its current reported value with large uncertainties ($R_{p} = 2.03_{-0.49}^{+0.61} R_{J}$), and thus is not as favourable for emission spectroscopy as its face-value ESM would imply.

We also cross-correlated models of SiO, CH$_{4}$, HCN, C$_{2}$H$_{2}$ and CO$_{2}$, but no significant detections were recorded (Appendix Fig. \ref{fig:no_detect_ccf}). 

\subsection{Retrieval results} \label{sec:retrievalres}


We obtain bounded abundance constraints for H$_{2}$O and CO, respectively finding values of log(H$_{2}$O) = $-4.42\pm{0.18}$ and log(CO) = $-3.85^{+0.33}_{-0.36}$.
Both retrieved abundances are under the solar model prediction values. We also recover upper limits on the VMR of SiO, CO$_{2}$, CH$_{4}$, HCN, C$_{2}$H$_{2}$ (see Figure \ref{fig:mix_ratio} and Table \ref{tab:results}). We note that a peak is present in the CH$_4$ posterior at the abundance expected for a solar-like composition in chemical equilibrium and also a positive correlation is found in its $K_{p}-V_{\mathrm{sys}}$ (Appendix Fig. \ref{fig:no_detect_ccf}). 
However, given that the cross-correlation signal is not significant, and its abundance posterior in the retrieval is not bounded, we do not consider this a detection.  Still, this may warrant further observations of \planet.  CH$_4$ was previously claimed to be detected in the dayside of the hot Jupiter HD 102195b using high-resolution spectroscopy \citep{guilluy_exoplanet_2019}, but a re-analysis of the same data with an updated CH$_4$ line list no longer showed any signal \citep{gandhi_molecular_2020}.

Our results do not favour grey clouds above the bar pressure level, likely indicating that \planet\ is not dominated by high-altitude clouds on its dayside.

\begin{figure*}
    \centering
    \includegraphics[width=\linewidth]{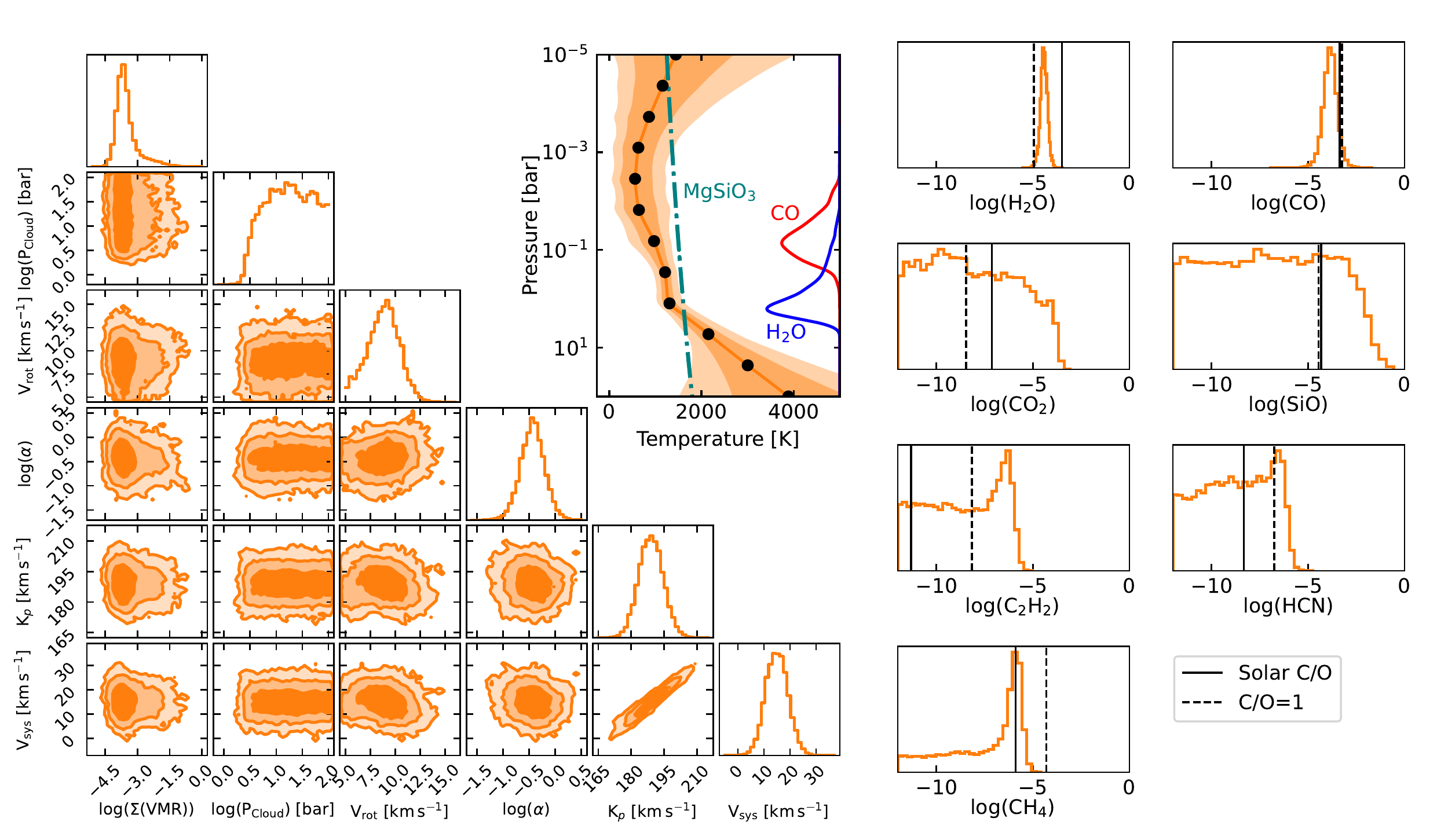}
    \caption{Summary of the results from the free chemistry retrieval of our \planet\ IGRINS dayside data set. The left corner plot includes the posteriors of planetary and orbital parameters. log($\Sigma$(VMR)) is the sum of the Volume Mixing Ratios (VMR) of all the retrieved species (shown separately for clarity). 
    log(P$_{\mathrm{Cloud}}$) is the (base 10) logarithm of the cloud-top pressure in units of bars.
    $V_{\mathrm{rot}}$ is the retrieved rotational broadening of \planet. log($\alpha$) is the model scaling factor used in the likelihood (Eq.\ \ref{eq:likelihood}). $K_{p}$ and $V_{\mathrm{sys}}$ are the orbital parameters as described in Eq.\ \ref{eq:dopplershift}.
    The top middle panel is the retrieved vertical non-inverted temperature profile of \planet\ with a smoothing prior $\sigma_{s} = 350$\,K\,dex$^{-2}$. The orange line is the median temperature profile while the darker and lighter orange shaded areas are the 1$\,\sigma$ (68\%) and 2$\,\sigma$ (95\%) regions, respectively. The 12 black points the pressure levels where temperature points are retrieved. The dark green dash-dotted line is the condensation curve of MgSiO$_3$. The blue and red curves represent the contribution functions for H$_2$O and CO, respectively. The H$_2$O probes around the 1 bar level, while the CO probes higher, around the 0.1 bar level.
    The right panels are the retrieved posteriors of the VMR of the fitted chemical species. The vertical lines represents the predicted thermochemical equilibrium abundance at 0.3 bar for a solar metallicity atmosphere with a uniform temperature profile of 1411\,K (see Figure \ref{fig:mix_ratio}). 
    The solid lines are from a model with solar C/O and the dashed lines are from a model with a C/O of 1.
    H$_{2}$O and CO are well constrained while only upper limits are inferred on other molecules. 
    The retrieved CO VMR is slightly below but consistent with the chemical equilibrium model. The H$_2$O abundance lies between the solar C/O and C/O=1 model predictions.
    }
    \label{fig:retrievalres}
\end{figure*}

We retrieved a non-inverted temperature profile (Figure \ref{fig:retrievalres}, middle panel).
This is expected for planets in this temperature range \citep[e.g.,][]{mansfield_unique_2021, baxter_transition_2020}.
The data probe a wide range of pressures in the mid atmosphere of the planet, where the absolute temperature gradient is high and the atmosphere is not too thick or thin (Figure \ref{fig:retrievalres}, middle panel red and blue curves). Due to this, the temperature profile is better constrained at those pressure levels. This causes the retrieved temperature at the top and bottom of the atmosphere to be unconstrained and the median gravitates towards the middle of the prior. This is what causes the seemingly inverted profile at the top of the atmosphere. Furthermore, our temperature prior prefers a profile with a lower curvature, explaining the straight profile at the bottom of the atmosphere where the data has little-to-no sensitivity.

We infer the model scaling factor to be log($\alpha$) = $-0.44\pm{0.21}$.  Although log($\alpha$) = $0$ (or $\alpha$ = $1$) is still within the retrieved 3\,$\sigma$ bounds, this value is low, suggesting that our model is overestimating the line contrast of the true signal.  In order to verify this, we ran a retrieval where we fixed the scaling factor to 1, to see what impact this had on the other retrieved parameters.  We found that this mainly only affected the retrieved thermal structure, where fixing $\alpha=1$ resulted in a smaller lapse rate at the pressures probed, with the retrieved temperature being similar in the upper atmosphere but slightly colder in the lower atmosphere. The interplay between temperature gradient and scaling factor can be understood as two competing effect to affect the line depth of spectral features in the model.  In this case, increasing $\alpha$ from its previously retrieved value (by setting it to 1) is now compensated by a lower temperature gradient, acting to similarly decrease the depth of spectral lines in the model.

The retrieved rotational broadening of \planet\ ($V_{\mathrm{rot}} = 8.9^{+1.5}_{-1.7}$ km\,s$^{-1}$) is slightly lower than the tidally-locked equatorial rotational broadening ($\sim$10.5\,km\,s$^{-1}$, assuming a radius of 2.03\,R$_J$).  While the rotation rate of the planet should be a lower-limit of the observed line broadening, with dynamical effect acting on top of this, finding a rotational broadening below tidally-locked equatorial rate is unexpected.  For example, recent dayside observations of the hot Jupiter WASP-43b using CRIRES$^+$ found a rotational broadening larger than expected \citep{lesjak_retrieval_2023}. They attributed this high broadening to a strong equatorial jet of several km$\,$s$^{-1}$. As \planet\ has a similar equilibrium temperature than WASP-43b, it might have similar atmospherical dynamics and would make this scenario probable.

\begin{center}
\begin{table}
\centering
\caption{\planet\ Retrieval Results}
\begin{tabular}{| c | c |}
 \hline
 \multicolumn{2}{|c|}{\textbf{Free chemistry retrieval}} \\
 \hline
 log(H$_{2}$O) & $-4.42\pm{0.18}$ \\
 log(CO) & $-3.85^{+0.33}_{-0.36}$ \\
 log(SiO) & $< -1.11$ ($3\,\sigma$ upper limit) \\
 log(CO$_{2}$) & $< -3.64$ ($3\,\sigma$ upper limit) \\
 log(CH$_{4}$) & $< -5.15$ ($3\,\sigma$ upper limit) \\
 log(HCN) & $< -5.77$ ($3\,\sigma$ upper limit) \\
 log(C$_{2}$H$_{2}$) & $< -5.68$ ($3\,\sigma$ upper limit) \\
 P$_{\mathrm{Cloud}}$ & $> 2.18$ bar ($3\,\sigma$ lower limit) \\
log($\alpha$) & $-0.44\pm{0.21}$ \\
 $V_{\mathrm{rot}}$ & $8.9^{+1.5}_{-1.7}$ km\,s$^{-1}$ \\
 $K_{p}$ & $188.6^{+5.6}_{-5.4}$ km\,s$^{-1}$ \\
 $V_{\mathrm{sys}}$ & $14.6^{+4.3}_{-4.1}$ km\,s$^{-1}$ \\
 \hline
 C/O (with SiO free) & $0.71^{+0.16}_{-0.47}$ \\
 C/O (SiO $< $ 3$\times$\,solar) 
 & $0.77^{+0.12}_{-0.20}$ \\
 $[$C/H$]$ & $-0.65^{+0.33}_{-0.35}$ \\
 $[$O/H$]$ & $-0.66^{+0.54}_{-0.31}$ \\
 \hline
 \multicolumn{2}{|c|}{\textbf{Chemical equilibrium retrieval}} \\
 \hline
 C/O & $0.72^{+0.13}_{-0.30}$ \\
 Metallicity, [M/H] & $-0.76^{+0.42}_{-0.48}$ \\
 \hline
\end{tabular}
\label{tab:results}
\end{table}
\end{center}

One hypothesis that could explain the relatively low retrieved $V_{\mathrm{rot}}$ value would be if \planet\ had a localized hot spot near the sub-stellar point.
The retrieved temperatures should be a weighted average of the observed projected face of the planet. Therefore, if the planet has a hot spot in its atmosphere \citep[e.g.,][]{coulombe_broadband_2023}, the observed flux, being proportional to temperature to the fourth power, would come disproportionately from this hotter region. In this scenario, the retrieved temperature profile, and all the retrieved molecular abundances, would be driven by the atmospheric conditions in this higher temperature but smaller-in-size region.
This would act to reduces the effective (flux-weighted) radius of \planet\ and could explain the smaller rotational broadening retrieved.

Another possibility is that \planet\ might simply have a radius smaller than the 2.03\,R$_J$ used in this analysis.
Without dynamical processes in \planet's atmosphere, the retrieved rotational broadening of $8.9$ km\,s$^{-1}$ would indicate that \planet's radius would be 1.72\,R$_J$, which is still within 1$\sigma$ of the radius recovered from transit \citep{nielsen_three_2020}.
With the relatively poor detection significance of molecular species given the ESM, and the fact that \planet\ is an outlier in the radius-mass space \citep[][see their Figure 10]{nielsen_three_2020}, it is likely that the radius is even smaller than 1.72\,R$_J$. In this scenario, the retrieved rotational broadening would be higher than the tidally-locked rotational broadening, likely due to additional atmospheric dynamical effects.

In the retrieval, we observe correlations between the sum of the volume mixing ratios of the retrieved species (log($\Sigma$(VMR))), the cloud-top pressure P$_{\mathrm{Cloud}}$, the model scaling factor (log($\alpha$)), the temperature structure, and the rotational broadening ($V_{\mathrm{rot}}$) (Figure \ref{fig:retrievalres}, left corner plot). These degeneracies arise from the fact that these parameters all affect the modelled line depths and shapes, stressing the importance of fitting these simultaneously.

We retrieved the Keplerian and systemic velocities of $188.6^{+5.6}_{-5.4}$ km\,s$^{-1}$ and $14.6^{+4.3}_{-4.1}$ km\,s$^{-1}$, respectively. 
$K_p$ is slightly lower than the expected value \citep[192.7 km\,s$^{-1}$, calculated with planetary parameters given in][]{nielsen_three_2020}, but still consistent within 1\,$\sigma$. 
The retrieved $V_{\mathrm{sys}}$ is blueshifted relative to the literature value \citep[21.04 km\,s$^{-1}$,][]{gaia_collaboration_vizier_2018}, but remains consistent at the 2\,$\sigma$ sigma level.

From retrieval results, we derive the ratio of carbon and oxygen atoms in the gas-phase of \planet's atmosphere.
The C/O is calculated using the following formula:
\begin{equation} \label{eq:ctoo}
    \mathrm{C/O} = \frac{n_{\mathrm{CO}} + n_{\mathrm{CO_2}} + n_{\mathrm{CH_4}} + n_{\mathrm{HCN}} + 2n_{\mathrm{C_2H_2}}}{n_{\mathrm{H_2O}} + n_{\mathrm{CO}} + 2n_{\mathrm{CO_2}} + n_{\mathrm{SiO}}}.
\end{equation}

We note that while we include SiO here because it can be an important oxygen-bearing molecule in hot Jupiter atmospheres (Figure \ref{fig:mix_ratio}), its abundance is poorly constrained by the retrieval due to SiO having no significant spectral features over the IGRINS wavelength range \citep{yurchenko_exomol_2022}.
Including SiO, the calculated gas-phase C/O is $0.71^{+0.16}_{-0.47}$, with the relatively large lower uncertainty due to SiO-rich compositions that cannot be ruled out from the IGRINS data alone.
If we instead omit scenarios where the SiO abundance is unrealistically high (greater than three times solar, or approximately 0.015\% of the planetary atmosphere, Figure \ref{fig:mix_ratio}), we measure an atmospheric C/O of $0.77^{+0.12}_{-0.20}$, now mostly driven by CO and H$_2$O as the main carbon- and oxygen-bearing molecules in the gas phase on \planet. This cutoff ensures that SiO is still included in the C/O determination, while excluding unlikely scenarios where 
both CO and H$_2$O are sub-solar but SiO would be super-solar.
However, we note that Si-rich atmospheric scenarios are possible under certain formation conditions, which would naturally give rise to atmospheres that are H$_2$O-poor due to a significant amount of oxygen atoms being bound in SiO \citep{chachan_breaking_2023}.



A possible explanation for the elevated C/O is that the oxygen is undetectable as it may be condensed out of the gas phase. Oxygen-bearing compounds such as MgSiO$_3$ may form in the atmosphere of \planet\ acting as a sink for oxygen atoms. We find that the temperature in the mid-atmosphere is lower than the condensation temperature of MgSiO$_3$ (Figure \ref{fig:retrievalres}, middle panel), indicating that condensation of that species is possible. This mechanism may even be stronger on the colder nightside, essentially trapping the oxygen in liquid or solid form.

We further calculate elemental abundance normalized to solar ([C/H] and [O/H]) defined as
\begin{equation} \label{eq:ctoh}
    \mathrm{[C/H]} = \log_{10} \left( \frac{n_{\mathrm{CO}} + n_{\mathrm{CO_2}} + n_{\mathrm{CH_4}} + n_{\mathrm{HCN}} + 2n_{\mathrm{C_2H_2}}}{2n_{\mathrm{H_2}} (n_{\mathrm{C_\odot}}/n_{\mathrm{H_\odot}})} \right)
\end{equation}
and
\begin{equation} \label{eq:otoh}
    \mathrm{[O/H]} = \log_{10} \left( \frac{n_{\mathrm{H_2O}} + n_{\mathrm{CO}} + 2n_{\mathrm{CO_2}} + n_{\mathrm{SiO}}}{2n_{\mathrm{H_2}} (n_{\mathrm{O_\odot}}/n_{\mathrm{H_\odot}})} \right).
\end{equation}

The derived gas-phase [C/H] and [O/H] are $-0.65^{+0.33}_{-0.35}$ and $-0.66^{+0.54}_{-0.31}$, respectively. This indicates a sub-solar metallicity, consistent within 2\,$\sigma$ of the solar values. The [Fe/H] of the host star HIP~65A is slightly super-solar \citep{nielsen_three_2020}, therefore \planet's metal enrichment can also be considered to be sub-stellar.

To verify our results, we also ran chemically consistent retrievals, fitting the C/O and overall metallicity. Under chemical equilibrium assumptions, the retrieved C/O is $0.72^{+0.13}_{-0.30}$ and the metallicity is $-0.76^{+0.42}_{-0.48}$, consistent with the values obtained from the free retrieval which makes no assumption on the chemistry but uses uniform-with-altitude abundance profiles.



\section{Discussion} \label{sec:discussion}

\subsection{Implication for \planet's formation} \label{sec:formation}

The present-day envelope composition of giant planets if the results of their evolution history. Retrieving atmospheric elemental abundance ratios such as the carbon-to-oxygen ratio and the metallicity are thus essential in the analysis of the giant planet formation. As envelope accretion occurs in a protoplanetary disk with a varying solid and gas composition as a function of the distance from the host star, the formation location of a planet may then determine its present-day atmospheric composition \citep[e.g.,][]{oberg_effects_2011, madhusudhan_toward_2014, oberg_excess_2016, booth_chemical_2017}. 
We can thus use the elemental ratios calculated from our retrieval results to speculate about what formation and migration scenarios may have given rise to \planet's atmospheric composition.

If we assume that the measured abundances in \planet's atmosphere are representative of the bulk envelope, one scenario that would naturally result in a sub-solar metallicity and somewhat super-solar C/O is if \planet\ was formed beyond the H$_{2}$O or CO$_2$ snowlines \citep{oberg_effects_2011}. The protoplanetary gas-phase metal enrichment drops with orbital separation as molecules condense into ices in the cold outer regions. Therefore, planets undergoing runaway gas accretion further out in the disc should have sub-solar metallicity with super-solar C/O, as inferred here.  \planet\ would then have migrated to its current orbital location.
The disk-free migration is the most probable migration scenario \citep{madhusudhan_toward_2014} as \planet\ would accrete few metals, and hence maintain its primordial sub-solar metallicity.

Alternatively, if a significant portion of metals are held in condensates (e.g., MgSiO$_3$) or in the interior \citep{thorngren_connecting_2019}, it is possible that \planet\ has a bulk content that is closer to solar/stellar.  In this sense, our measurement of the C/H and O/H in the atmosphere serves as a lower-limit of the bulk envelope composition.  If this is the case, formation scenarios inwards of the water snowline and/or via gravitational instability can also not be ruled-out.

\subsection{Benchmarking our HRCCS pipeline} \label{sec:W77}

\begin{figure}
    \centering
    \includegraphics[width=\linewidth]{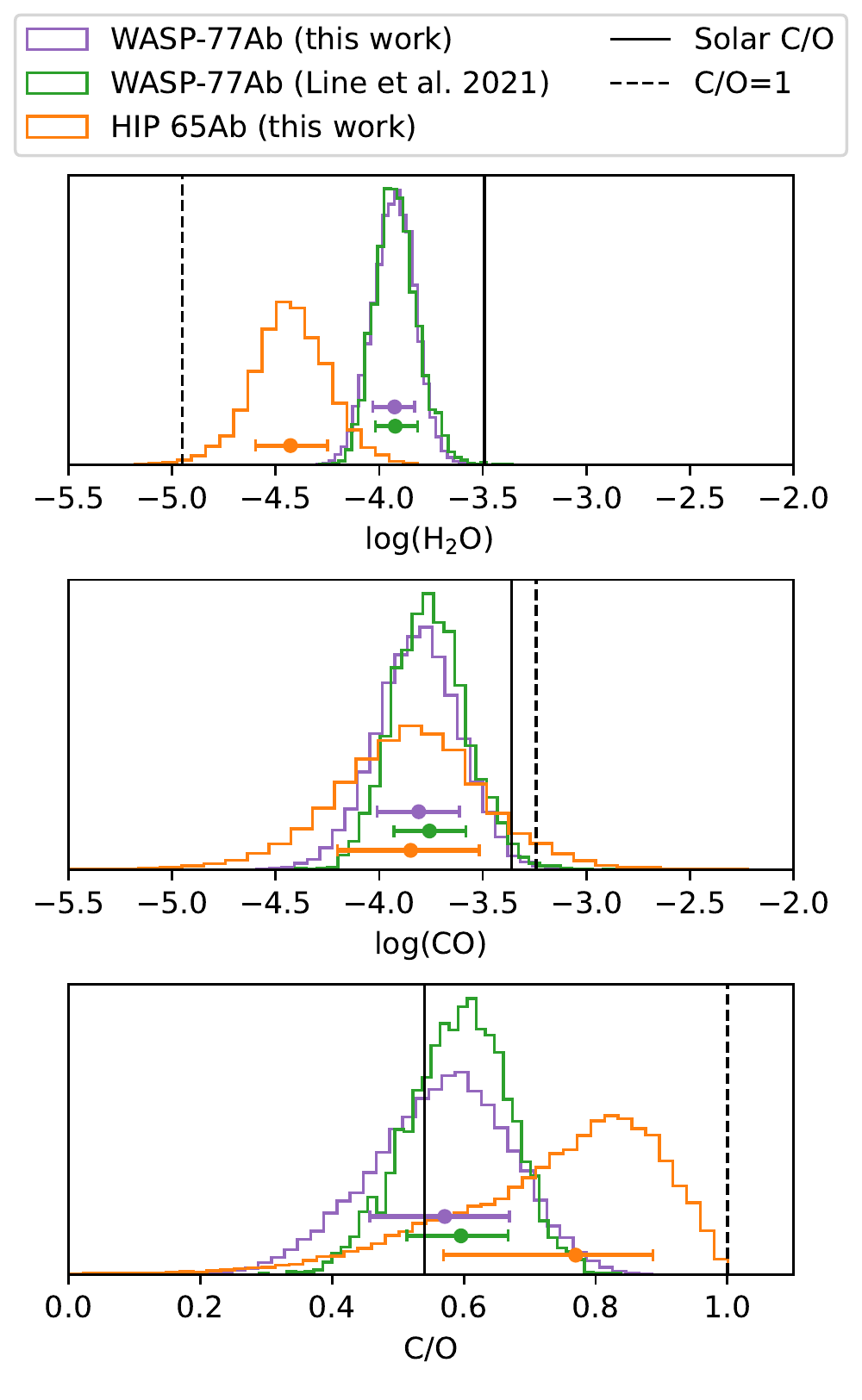}
    \caption{
    Benchmarking of our analysis framework and comparison to WASP-77Ab.
    Here we show the H$_{2}$O and CO abundances, and C/O retrieved for \planet\ compared to IGRINS emission observations of WASP-77Ab from \citet{line_solar_2021}.  Our analysis of the WASP-77Ab data (purple) well-reproduces the molecular abundances and C/O reported by \citet{line_solar_2021} (green). In contrast, using our pipeline, \planet\ (orange) shows a comparatively depleted H$_{2}$O abundance and higher C/O. 
    Here we show the retrieved C/O from the main free chemistry retrieval with high SiO scenarios removed.
    The black lines represent the abundances taken at 0.3 bar of a solar metallicity model (solid is C/O = solar, dashed is C/O = 1) assuming an isothermal 1411\,K atmosphere (Figure \ref{fig:mix_ratio}). 
    }
    \label{fig:HIP65W77VMR}
\end{figure}

In order to benchmark our analysis, we tested our high-resolution cross-correlation spectroscopy (HRCCS) pipeline by recreating established literature results. 
We opted to recreate a small-scale version of the analysis presented in \citet{line_solar_2021}, similar to what they have done with the HyDRA-H code \citep{gandhi_hydra-h_2019}.  
We used public data of WASP-77Ab taken with IGRINS on 14 December 2020. WASP-77Ab is a hot Jupiter ($T_{\mathrm{eq}} = 1740$\,K) in a 1.36 day period \citep{maxted_wasp-77_2012}. It has a radius of $1.21 R_{J}$ and a mass of $1.76 M_{J}$. We refer the reader to \citet{line_solar_2021} for more information on their analysis and results.

For our retrieval of WASP-77Ab, we only fitted the abundance of H$_2$O and CO and fixed the model scaling factor to $\log(\alpha) = 0$.
We obtain retrieval results that are fully consistent with \citet{line_solar_2021}. From our analysis of the WASP-77Ab IGRINS data we measure abundances of log(H$_{2}$O) = $-3.93^{+0.10}_{-0.11}$ and log(CO) = $-3.81\pm{0.20}$. These values are extremely close with \citet{line_solar_2021} who retrieved log(H$_{2}$O) = $-3.93^{+0.10}_{-0.09}$ and log(CO) = $-3.77^{+0.18}_{-0.16}$ (Figure \ref{fig:HIP65W77VMR}). This is encouraging as we are using a fully independent HRCCS reduction pipeline \citep{pelletier_where_2021}, modelling framework \citep{benneke_strict_2015}, and likelihood prescription \citep{gibson_detection_2020}, while still producing almost identical results. Our results are also consistent with the HyDRA-H framework retrieval presented in \citet{line_solar_2021}. This test serves as a useful benchmark of our pipeline, and also serves as a further validation of the results reported in \citet{line_solar_2021}.

With both planets analysed using our framework, we note differences between the retrieved abundances of WASP-77Ab and \planet\ (Figure \ref{fig:HIP65W77VMR}). The retrieved \planet\ CO abundance is similar to that of WASP-77Ab. \planet's H$_2$O content, however, is lower compared to WASP-77Ab. Its VMR is about than 0.5 dex lower than WASP-77Ab, illustrating again the depletion of H$_2$O in \planet's atmosphere. This difference causes the overall higher observed C/O on \planet. 
A possible explanation to explain \planet's lower water abundance relative to WASP-77Ab is if oxygen is further condensed out of the gas-phase (in species such as MgSiO$_3$) in \planet\ than in WASP-77Ab. \planet\ ($T_{\mathrm{eq}} = 1411$\,K) has a lower temperature than WASP-77Ab ($T_{\mathrm{eq}} = 1740$\,K), so an oxygen-bearing molecule with condensation temperature between the temperatures of the two planets could explain the difference in the measured gas-phase oxygen content between them, while still having the same global oxygen content.

\section{Conclusion} \label{sec:conclusion}

We analysed 3.5 hours of IGRINS dayside emission spectroscopic data from the hot Jupiter \planet. Using a Bayesian analysis approach, we constrained the water and carbon monoxide abundance in its atmosphere to be log(H$_{2}$O) = $-4.42\pm{0.18}$ and log(CO) = $-3.85^{+0.33}_{-0.36}$. Other carbon- and oxygen-bearing molecules were not detected significantly, with only upper limits obtained on their abundances. Our results suggest that \planet\ has a sub-solar metallicity, with a slightly elevated super-solar C/O. The measured composition is likely either the result of \planet's formation and evolution history, or the result of missing metals being condensed out of the gas phase of the atmosphere.

The retrieved temperatures reveals a non-inverted temperature structure in \planet's atmosphere. This profile is consistent with expectations for a hot Jupiter of \planet's equilibrium temperature. Our results favour a dayside atmosphere for \planet\ that is not dominated by high-altitude clouds. $K_{p}$ and $V_{\mathrm{sys}}$ are retrieved at values near the expected literature values. We retrieved a rotational broadening smaller than the tidally-locked equatorial speed of \planet, and also recover a low scaling factor.  These findings, combined with the relatively weak cross-correlation detections compared to similar IGRINS observations of WASP-77Ab strongly suggests the \planet\ has a radius on the significant lower end of its reported values of $R_{p} = 2.03_{-0.49}^{+0.61} R_{J}$.



From our retrieved molecule volume mixing ratios, we derived elemental abundance ratios, finding \planet\ to have C/H and O/H values that are sub-solar, with a slightly elevated C/O ratio. Such an atmospheric composition is consistent with \planet\ having formed further out in the protoplanetary disc, before migrating to its present-day orbital position.  This likely would have occuved via disc-free migration mechanisms in order to not accrete a high amount of planetesimals in order to preserve its sub-solar metallicity.

Finally, we benchmark our analysis framework on IGRINS data of WASP-77Ab \citep{line_solar_2021}, finding that we are able to obtain fully consistent results despite our fully independent detrending and atmospheric modelling pipelines.



\section{Acknowledgments}
We thank the reviewer for providing comments which improved the quality of this manuscript.
This work used The Immersion Grating Infrared Spectrometer (IGRINS) that was developed under a collaboration between the University of Texas at Austin and the Korea Astronomy and Space Science Institute (KASI) with the financial support of the US National Science Foundation under grants AST-1229522, AST-1702267 and AST-1908892, McDonald Observatory of the University of Texas at Austin, the Korean GMT Project of KASI, the Mt.\ Cuba Astronomical Foundation and Gemini Observatory. This work is based on observations obtained at the international Gemini Observatory, a program of NSF’s NOIRLab, which is managed by the Association of Universities for Research in Astronomy (AURA) under a cooperative agreement with the National Science Foundation on behalf of the Gemini Observatory partnership: the National Science Foundation (United States), National Research Council (Canada), Agencia Nacional de Investigación y Desarrollo (Chile), Ministerio de Ciencia, Tecnología e Innovación (Argentina), Ministério da Ciência, Tecnologia, Inovações e Comunicações (Brazil), and Korea Astronomy and Space Science Institute (Republic of Korea).
L.B.\ acknowledges funding by the Natural Sciences and Engineering Research Council of Canada (NSERC).
S.P.\ is partially supported by the Technologies for Exo-Planetary Science (TEPS) NSERC CREATE Trainee Program.
B.B.\ acknowledges funding by the NSERC, and the Fonds de Recherche du Québec -- Nature et Technologies (FRQNT).
This project has been carried out within the framework of the National Centre of Competence in Research PlanetS supported by the Swiss National Science Foundation under grant 51NF40\_205606. The authors acknowledge the financial support of the SNSF.
We are thankful to Michael Line for making the WASP-77Ab IGRINS data and associated data reduction codes publicly available.

%

\vspace{5mm}
\facility{Gemini-South (IGRINS)}


\software{Astropy \citep{robitaille_astropy_2013, the_astropy_collaboration_astropy_2018, the_astropy_collaboration_astropy_2022},
          NumPy \citep{harris_array_2020}, 
          SciPy \citep{virtanen_scipy_2020},
          Matplotlib \citep{hunter_matplotlib_2007},
          emcee \citep{foreman-mackey_emcee_2013}, 
          corner \citep{foreman-mackey_cornerpy_2016}
          }



\appendix

$K_{p}-V_{\mathrm{sys}}$ maps for other notable carbon- and oxygen-bearing molecules.

\begin{figure*}
    \centering
    \includegraphics[width=\linewidth]{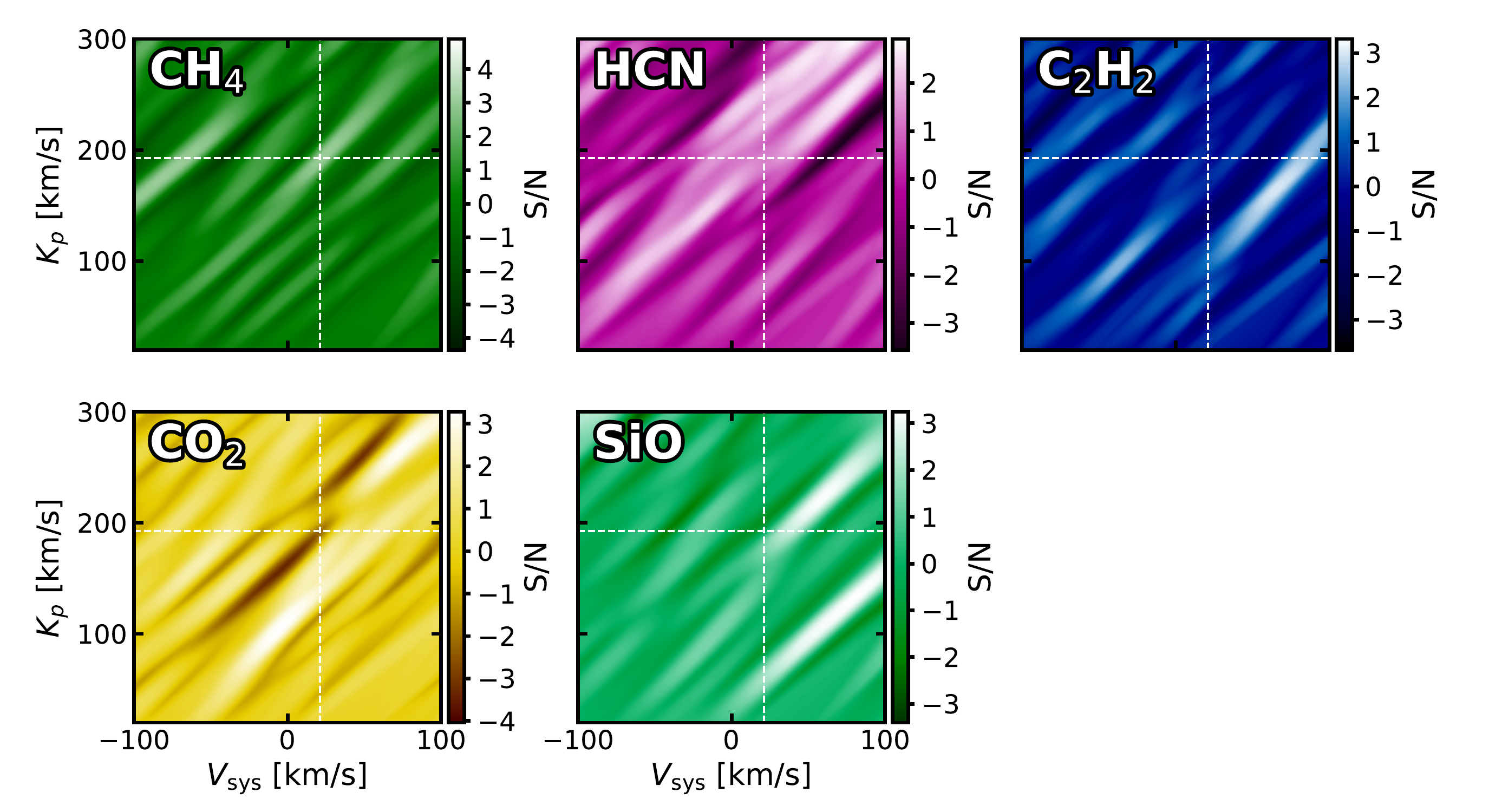}
    \caption{$K_{p}-V_{\mathrm{sys}}$ cross-correlation maps (same as Figure \ref{fig:detect_ccfs}), but for molecules showing no significant detections near the expected planetary position. Despite these non-detections, we still fit these species in the retrieval to derive upper limits on their abundances.}
    \label{fig:no_detect_ccf}
\end{figure*}




\bibliography{main}{}

\begin{thebibliography}{}
\expandafter\ifx\csname natexlab\endcsname\relax\def\natexlab#1{#1}\fi
\providecommand{\url}[1]{\href{#1}{#1}}
\providecommand{\dodoi}[1]{doi:~\href{http://doi.org/#1}{\nolinkurl{#1}}}
\providecommand{\doeprint}[1]{\href{http://ascl.net/#1}{\nolinkurl{http://ascl.net/#1}}}
\providecommand{\doarXiv}[1]{\href{https://arxiv.org/abs/#1}{\nolinkurl{https://arxiv.org/abs/#1}}}

\bibitem[{Barber {et~al.}(2014)Barber, Strange, Hill, Polyansky, Mellau,
  Yurchenko, \& Tennyson}]{barber_exomol_2014}
Barber, R.~J., Strange, J.~K., Hill, C., {et~al.} 2014, Monthly Notices of the
  Royal Astronomical Society, 437, 1828, \dodoi{10.1093/mnras/stt2011}

\bibitem[{Barber {et~al.}(2006)Barber, Tennyson, Harris, \&
  Tolchenov}]{barber_high-accuracy_2006}
Barber, R.~J., Tennyson, J., Harris, G.~J., \& Tolchenov, R.~N. 2006, Monthly
  Notices of the Royal Astronomical Society, 368, 1087,
  \dodoi{10.1111/j.1365-2966.2006.10184.x}

\bibitem[{Baxter {et~al.}(2020)Baxter, Désert, Parmentier, Line, Fortney,
  Arcangeli, Bean, Todorov, \& Mansfield}]{baxter_transition_2020}
Baxter, C., Désert, J.-M., Parmentier, V., {et~al.} 2020, Astronomy \&
  Astrophysics, 639, A36, \dodoi{10.1051/0004-6361/201937394}

\bibitem[{Benneke(2015)}]{benneke_strict_2015}
Benneke, B. 2015, Strict {Upper} {Limits} on the {Carbon}-to-{Oxygen} {Ratios}
  of {Eight} {Hot} {Jupiters} from {Self}-{Consistent} {Atmospheric}
  {Retrieval},  arXiv, \dodoi{10.48550/arXiv.1504.07655}

\bibitem[{Benneke \& Seager(2012)}]{benneke_atmospheric_2012}
Benneke, B., \& Seager, S. 2012, The Astrophysical Journal, 753, 100,
  \dodoi{10.1088/0004-637X/753/2/100}

\bibitem[{Benneke \& Seager(2013)}]{benneke_how_2013}
---. 2013, The Astrophysical Journal, 778, 153,
  \dodoi{10.1088/0004-637X/778/2/153}

\bibitem[{Benneke {et~al.}(2019{\natexlab{a}})Benneke, Wong, Piaulet, Knutson,
  Lothringer, Morley, Crossfield, Gao, Greene, Dressing, Dragomir, Howard,
  McCullough, Kempton, Fortney, \& Fraine}]{benneke_water_2019}
Benneke, B., Wong, I., Piaulet, C., {et~al.} 2019{\natexlab{a}}, The
  Astrophysical Journal Letters, 887, L14, \dodoi{10.3847/2041-8213/ab59dc}

\bibitem[{Benneke {et~al.}(2019{\natexlab{b}})Benneke, Knutson, Lothringer,
  Crossfield, Moses, Morley, Kreidberg, Fulton, Dragomir, Howard, Wong,
  Désert, McCullough, Kempton, Fortney, Gilliland, Deming, \&
  Kammer}]{benneke_sub-neptune_2019}
Benneke, B., Knutson, H.~A., Lothringer, J., {et~al.} 2019{\natexlab{b}},
  Nature Astronomy, 3, 813, \dodoi{10.1038/s41550-019-0800-5}

\bibitem[{Birkby(2018)}]{birkby_spectroscopic_2018}
Birkby, J.~L. 2018, in Handbook of {Exoplanets}, ed. H.~J. Deeg \& J.~A.
  Belmonte (Cham: Springer International Publishing), 1485--1508,
  \dodoi{10.1007/978-3-319-55333-7_16}

\bibitem[{Birkby {et~al.}(2017)Birkby, Kok, Brogi, Schwarz, \&
  Snellen}]{birkby_discovery_2017}
Birkby, J.~L., Kok, R. J.~d., Brogi, M., Schwarz, H., \& Snellen, I. A.~G.
  2017, The Astronomical Journal, 153, 138, \dodoi{10.3847/1538-3881/aa5c87}

\bibitem[{Booth {et~al.}(2017)Booth, Clarke, Madhusudhan, \&
  Ilee}]{booth_chemical_2017}
Booth, R.~A., Clarke, C.~J., Madhusudhan, N., \& Ilee, J.~D. 2017, Monthly
  Notices of the Royal Astronomical Society, 469, 3994,
  \dodoi{10.1093/mnras/stx1103}

\bibitem[{Borysow(2002)}]{borysow_collision-induced_2002}
Borysow, A. 2002, Astronomy \& Astrophysics, 390, 779,
  \dodoi{10.1051/0004-6361:20020555}

\bibitem[{Brogi \& Line(2019)}]{brogi_retrieving_2019}
Brogi, M., \& Line, M.~R. 2019, The Astronomical Journal, 157, 114,
  \dodoi{10.3847/1538-3881/aaffd3}

\bibitem[{Brogi {et~al.}(2012)Brogi, Snellen, de~Kok, Albrecht, Birkby, \&
  de~Mooij}]{brogi_signature_2012}
Brogi, M., Snellen, I. A.~G., de~Kok, R.~J., {et~al.} 2012, Nature, 486, 502,
  \dodoi{10.1038/nature11161}

\bibitem[{Brogi {et~al.}(2023)Brogi, Emeka-Okafor, Line, Gandhi, Pino, Kempton,
  Rauscher, Parmentier, Bean, Mace, Cowan, Shkolnik, Wardenier, Mansfield,
  Welbanks, Smith, Fortney, Birkby, Zalesky, Dang, Patience, \&
  Désert}]{brogi_roasting_2023}
Brogi, M., Emeka-Okafor, V., Line, M.~R., {et~al.} 2023, The Astronomical
  Journal, 165, 91, \dodoi{10.3847/1538-3881/acaf5c}

\bibitem[{Cabot {et~al.}(2019)Cabot, Madhusudhan, Hawker, \&
  Gandhi}]{cabot_robustness_2019}
Cabot, S. H.~C., Madhusudhan, N., Hawker, G.~A., \& Gandhi, S. 2019, Monthly
  Notices of the Royal Astronomical Society, 482, 4422,
  \dodoi{10.1093/mnras/sty2994}

\bibitem[{Chachan {et~al.}(2023)Chachan, Knutson, Lothringer, \&
  Blake}]{chachan_breaking_2023}
Chachan, Y., Knutson, H.~A., Lothringer, J., \& Blake, G.~A. 2023, The
  Astrophysical Journal, 943, 112, \dodoi{10.3847/1538-4357/aca614}

\bibitem[{Cheverall {et~al.}(2023)Cheverall, Madhusudhan, \&
  Holmberg}]{cheverall_robustness_2023}
Cheverall, C.~J., Madhusudhan, N., \& Holmberg, M. 2023, Monthly Notices of the
  Royal Astronomical Society, 522, 661, \dodoi{10.1093/mnras/stad648}

\bibitem[{Chubb {et~al.}(2020)Chubb, Tennyson, \&
  Yurchenko}]{chubb_exomol_2020}
Chubb, K.~L., Tennyson, J., \& Yurchenko, S.~N. 2020, Monthly Notices of the
  Royal Astronomical Society, 493, 1531, \dodoi{10.1093/mnras/staa229}

\bibitem[{Collaboration {et~al.}(2018)Collaboration, Price-Whelan, Sipőcz,
  Günther, Lim, Crawford, Conseil, Shupe, Craig, Dencheva, Ginsburg,
  VanderPlas, Bradley, Pérez-Suárez, de~Val-Borro, Aldcroft, Cruz,
  Robitaille, Tollerud, Ardelean, Babej, Bachetti, Bakanov, Bamford, Barentsen,
  Barmby, Baumbach, Berry, Biscani, Boquien, Bostroem, Bouma, Brammer, Bray,
  Breytenbach, Buddelmeijer, Burke, Calderone, Rodríguez, Cara, Cardoso,
  Cheedella, Copin, Crichton, DÁvella, Deil, Depagne, Dietrich, Donath,
  Droettboom, Earl, Erben, Fabbro, Ferreira, Finethy, Fox, Garrison, Gibbons,
  Goldstein, Gommers, Greco, Greenfield, Groener, Grollier, Hagen, Hirst,
  Homeier, Horton, Hosseinzadeh, Hu, Hunkeler, Ivezić, Jain, Jenness, Kanarek,
  Kendrew, Kern, Kerzendorf, Khvalko, King, Kirkby, Kulkarni, Kumar, Lee, Lenz,
  Littlefair, Ma, Macleod, Mastropietro, McCully, Montagnac, Morris, Mueller,
  Mumford, Muna, Murphy, Nelson, Nguyen, Ninan, Nöthe, Ogaz, Oh, Parejko,
  Parley, Pascual, Patil, Patil, Plunkett, Prochaska, Rastogi, Janga, Sabater,
  Sakurikar, Seifert, Sherbert, Sherwood-Taylor, Shih, Sick, Silbiger,
  Singanamalla, Singer, Sladen, Sooley, Sornarajah, Streicher, Teuben, Thomas,
  Tremblay, Turner, Terrón, van Kerkwijk, de~la Vega, Watkins, Weaver,
  Whitmore, Woillez, \& Zabalza}]{the_astropy_collaboration_astropy_2018}
Collaboration, T.~A., Price-Whelan, A.~M., Sipőcz, B.~M., {et~al.} 2018, The
  {Astropy} {Project}: {Building} an inclusive, open-science project and status
  of the v2.0 core package, \dodoi{10.3847/1538-3881/aabc4f}

\bibitem[{Collaboration {et~al.}(2022)Collaboration, Price-Whelan, Lim, Earl,
  Starkman, Bradley, Shupe, Patil, Corrales, Brasseur, Nöthe, Donath,
  Tollerud, Morris, Ginsburg, Vaher, Weaver, Tocknell, Jamieson, van Kerkwijk,
  Robitaille, Merry, Bachetti, Günther, Aldcroft, Alvarado-Montes, Archibald,
  Bódi, Bapat, Barentsen, Bazán, Biswas, Boquien, Burke, Cara, Cara, Conroy,
  Conseil, Craig, Cross, Cruz, D'Eugenio, Dencheva, Devillepoix, Dietrich,
  Eigenbrot, Erben, Ferreira, Foreman-Mackey, Fox, Freij, Garg, Geda, Glattly,
  Gondhalekar, Gordon, Grant, Greenfield, Groener, Guest, Gurovich, Handberg,
  Hart, Hatfield-Dodds, Homeier, Hosseinzadeh, Jenness, Jones, Joseph,
  Kalmbach, Karamehmetoglu, Kałuszyński, Kelley, Kern, Kerzendorf, Koch,
  Kulumani, Lee, Ly, Ma, MacBride, Maljaars, Muna, Murphy, Norman, O'Steen,
  Oman, Pacifici, Pascual, Pascual-Granado, Patil, Perren, Pickering, Rastogi,
  Roulston, Ryan, Rykoff, Sabater, Sakurikar, Salgado, Sanghi, Saunders,
  Savchenko, Schwardt, Seifert-Eckert, Shih, Jain, Shukla, Sick, Simpson,
  Singanamalla, Singer, Singhal, Sinha, Sipőcz, Spitler, Stansby, Streicher,
  Šumak, Swinbank, Taranu, Tewary, Tremblay, de~Val-Borro, Van~Kooten,
  Vasović, Verma, Cardoso, Williams, Wilson, Winkel, Wood-Vasey, Xue, Yoachim,
  ZHANG, \& Zonca}]{the_astropy_collaboration_astropy_2022}
Collaboration, T.~A., Price-Whelan, A.~M., Lim, P.~L., {et~al.} 2022, The
  {Astropy} {Project}: {Sustaining} and {Growing} a {Community}-oriented
  {Open}-source {Project} and the {Latest} {Major} {Release} (v5.0) of the
  {Core} {Package}, \dodoi{10.3847/1538-4357/ac7c74}

\bibitem[{Coulombe {et~al.}(2023)Coulombe, Benneke, Challener, Piette, Wiser,
  Mansfield, MacDonald, Beltz, Feinstein, Radica, Savel, Dos~Santos, Bean,
  Parmentier, Wong, Rauscher, Komacek, Kempton, Tan, Hammond, Lewis, Line, Lee,
  Shivkumar, Crossfield, Nixon, Rackham, Wakeford, Welbanks, Zhang, Batalha,
  Berta-Thompson, Changeat, Désert, Espinoza, Goyal, Harrington, Knutson,
  Kreidberg, López-Morales, Shporer, Sing, Stevenson, Aggarwal, Ahrer, Alam,
  Bell, Blecic, Caceres, Carter, Casewell, Crouzet, Cubillos, Decin, Fortney,
  Gibson, Heng, Henning, Iro, Kendrew, Lagage, Leconte, Lendl, Lothringer,
  Mancini, Mikal-Evans, Molaverdikhani, Nikolov, Ohno, Palle, Piaulet,
  Redfield, Roy, Tsai, Venot, \& Wheatley}]{coulombe_broadband_2023}
Coulombe, L.-P., Benneke, B., Challener, R., {et~al.} 2023, Nature, 620, 292,
  \dodoi{10.1038/s41586-023-06230-1}

\bibitem[{Foreman-Mackey(2016)}]{foreman-mackey_cornerpy_2016}
Foreman-Mackey, D. 2016, Journal of Open Source Software, 1, 24,
  \dodoi{10.21105/joss.00024}

\bibitem[{Foreman-Mackey {et~al.}(2013)Foreman-Mackey, Hogg, Lang, \&
  Goodman}]{foreman-mackey_emcee_2013}
Foreman-Mackey, D., Hogg, D.~W., Lang, D., \& Goodman, J. 2013, Publications of
  the Astronomical Society of the Pacific, 125, 306, \dodoi{10.1086/670067}

\bibitem[{{Gaia Collaboration}(2018)}]{gaia_collaboration_vizier_2018}
{Gaia Collaboration}. 2018, VizieR Online Data Catalog, I/345.
\newblock \url{https://ui.adsabs.harvard.edu/abs/2018yCat.1345....0G}

\bibitem[{Gandhi {et~al.}(2019)Gandhi, Madhusudhan, Hawker, \&
  Piette}]{gandhi_hydra-h_2019}
Gandhi, S., Madhusudhan, N., Hawker, G., \& Piette, A. 2019, The Astronomical
  Journal, 158, 228, \dodoi{10.3847/1538-3881/ab4efc}

\bibitem[{Gandhi {et~al.}(2020)Gandhi, Brogi, Yurchenko, Tennyson, Coles, Webb,
  Birkby, Guilluy, Hawker, Madhusudhan, Bonomo, \&
  Sozzetti}]{gandhi_molecular_2020}
Gandhi, S., Brogi, M., Yurchenko, S.~N., {et~al.} 2020, Monthly Notices of the
  Royal Astronomical Society, 495, 224, \dodoi{10.1093/mnras/staa981}

\bibitem[{Gibson {et~al.}(2019)Gibson, de Mooij, Evans, Merritt, Nikolov,
  Sing, \& Watson}]{gibson_revisiting_2019}
Gibson, N.~P., de Mooij, E. J.~W., Evans, T.~M., {et~al.} 2019, Monthly
  Notices of the Royal Astronomical Society, 482, 606,
  \dodoi{10.1093/mnras/sty2722}

\bibitem[{Gibson {et~al.}(2022)Gibson, Nugroho, Lothringer, Maguire, \&
  Sing}]{gibson_relative_2022}
Gibson, N.~P., Nugroho, S.~K., Lothringer, J., Maguire, C., \& Sing, D.~K.
  2022, Monthly Notices of the Royal Astronomical Society, 512, 4618,
  \dodoi{10.1093/mnras/stac091}

\bibitem[{Gibson {et~al.}(2020)Gibson, Merritt, Nugroho, Cubillos, de Mooij,
  Mikal-Evans, Fossati, Lothringer, Nikolov, Sing, Spake, Watson, \&
  Wilson}]{gibson_detection_2020}
Gibson, N.~P., Merritt, S., Nugroho, S.~K., {et~al.} 2020, Monthly Notices of
  the Royal Astronomical Society, 493, 2215, \dodoi{10.1093/mnras/staa228}

\bibitem[{Guilluy {et~al.}(2019)Guilluy, Sozzetti, Brogi, Bonomo, Giacobbe,
  Claudi, \& Benatti}]{guilluy_exoplanet_2019}
Guilluy, G., Sozzetti, A., Brogi, M., {et~al.} 2019, Astronomy \& Astrophysics,
  625, A107, \dodoi{10.1051/0004-6361/201834615}

\bibitem[{Hargreaves {et~al.}(2020)Hargreaves, Gordon, Rey, Nikitin, Tyuterev,
  Kochanov, \& Rothman}]{hargreaves_accurate_2020}
Hargreaves, R.~J., Gordon, I.~E., Rey, M., {et~al.} 2020, The Astrophysical
  Journal Supplement Series, 247, 55, \dodoi{10.3847/1538-4365/ab7a1a}

\bibitem[{Harris {et~al.}(2020)Harris, Millman, van~der Walt, Gommers,
  Virtanen, Cournapeau, Wieser, Taylor, Berg, Smith, Kern, Picus, Hoyer, van
  Kerkwijk, Brett, Haldane, del Río, Wiebe, Peterson, Gérard-Marchant,
  Sheppard, Reddy, Weckesser, Abbasi, Gohlke, \& Oliphant}]{harris_array_2020}
Harris, C.~R., Millman, K.~J., van~der Walt, S.~J., {et~al.} 2020, Nature, 585,
  357, \dodoi{10.1038/s41586-020-2649-2}

\bibitem[{Harris {et~al.}(2006)Harris, Tennyson, Kaminsky, Pavlenko, \&
  Jones}]{harris_improved_2006}
Harris, G.~J., Tennyson, J., Kaminsky, B.~M., Pavlenko, Y.~V., \& Jones, H.
  R.~A. 2006, Monthly Notices of the Royal Astronomical Society, 367, 400,
  \dodoi{10.1111/j.1365-2966.2005.09960.x}

\bibitem[{Holmberg \& Madhusudhan(2022)}]{holmberg_first_2022}
Holmberg, M., \& Madhusudhan, N. 2022, The Astronomical Journal, 164, 79,
  \dodoi{10.3847/1538-3881/ac77eb}

\bibitem[{Hunter(2007)}]{hunter_matplotlib_2007}
Hunter, J.~D. 2007, Computing in Science \& Engineering, 9, 90,
  \dodoi{10.1109/MCSE.2007.55}

\bibitem[{Kempton {et~al.}(2018)Kempton, Bean, Louie, Deming, Koll, Mansfield,
  Christiansen, López-Morales, Swain, Zellem, Ballard, Barclay, Barstow,
  Batalha, Beatty, Berta-Thompson, Birkby, Buchhave, Charbonneau, Cowan,
  Crossfield, Val-Borro, Doyon, Dragomir, Gaidos, Heng, Hu, Kane, Kreidberg,
  Mallonn, Morley, Narita, Nascimbeni, Pallé, Quintana, Rauscher, Seager,
  Shkolnik, Sing, Sozzetti, Stassun, Valenti, \&
  Essen}]{kempton_framework_2018}
Kempton, E. M.-R., Bean, J.~L., Louie, D.~R., {et~al.} 2018, Publications of
  the Astronomical Society of the Pacific, 130, 114401,
  \dodoi{10.1088/1538-3873/aadf6f}

\bibitem[{Knutson {et~al.}(2014)Knutson, Benneke, Deming, \&
  Homeier}]{knutson_featureless_2014}
Knutson, H.~A., Benneke, B., Deming, D., \& Homeier, D. 2014, Nature, 505, 66,
  \dodoi{10.1038/nature12887}

\bibitem[{Kreidberg {et~al.}(2014)Kreidberg, Bean, Désert, Benneke, Deming,
  Stevenson, Seager, Berta-Thompson, Seifahrt, \&
  Homeier}]{kreidberg_clouds_2014}
Kreidberg, L., Bean, J.~L., Désert, J.-M., {et~al.} 2014, Nature, 505, 69,
  \dodoi{10.1038/nature12888}

\bibitem[{Lee \& Gullikson(2016)}]{lee_plp_2016}
Lee, J.-J., \& Gullikson, K. 2016, plp: v2.1 alpha 3,  Zenodo,
  \dodoi{10.5281/zenodo.56067}

\bibitem[{Lesjak {et~al.}(2023)Lesjak, Nortmann, Yan, Cont, Reiners, Piskunov,
  Hatzes, Boldt-Christmas, Czesla, Heiter, Kochukhov, Lavail, Nagel, Rains,
  Rengel, Rodler, Seemann, \& Shulyak}]{lesjak_retrieval_2023}
Lesjak, F., Nortmann, L., Yan, F., {et~al.} 2023, Astronomy \& Astrophysics,
  678, A23, \dodoi{10.1051/0004-6361/202347151}

\bibitem[{Li {et~al.}(2015)Li, Gordon, Rothman, Tan, Hu, Kassi, Campargue, \&
  Medvedev}]{li_rovibrational_2015}
Li, G., Gordon, I.~E., Rothman, L.~S., {et~al.} 2015, The Astrophysical Journal
  Supplement Series, 216, 15, \dodoi{10.1088/0067-0049/216/1/15}

\bibitem[{Line {et~al.}(2021)Line, Brogi, Bean, Gandhi, Zalesky, Parmentier,
  Smith, Mace, Mansfield, Kempton, Fortney, Shkolnik, Patience, Rauscher,
  Désert, \& Wardenier}]{line_solar_2021}
Line, M.~R., Brogi, M., Bean, J.~L., {et~al.} 2021, Nature, 598, 580,
  \dodoi{10.1038/s41586-021-03912-6}

\bibitem[{Lothringer {et~al.}(2021)Lothringer, Rustamkulov, Sing, Gibson,
  Wilson, \& Schlaufman}]{lothringer_new_2021}
Lothringer, J.~D., Rustamkulov, Z., Sing, D.~K., {et~al.} 2021, The
  Astrophysical Journal, 914, 12, \dodoi{10.3847/1538-4357/abf8a9}

\bibitem[{Mace {et~al.}(2018)Mace, Sokal, Lee, Oh, Park, Lee, Good, MacQueen,
  Oh, Kaplan, Kidder, Chun, Yuk, Jeong, Pak, Kim, Nah, Lee, Yu, Hwang, Park,
  Kim, Chinn, Peck, Diaz, Rutten, Prato, Jacoby, Cornelius, Hardesty, DeGroff,
  Dunham, Levine, Nofi, Lopez-Valdivia, Weinberger, \&
  Jaffe}]{mace_igrins_2018}
Mace, G., Sokal, K., Lee, J.-J., {et~al.} 2018, in Ground-based and {Airborne}
  {Instrumentation} for {Astronomy} {VII}, Vol. 10702 (SPIE), 204--221,
  \dodoi{10.1117/12.2312345}

\bibitem[{Madhusudhan {et~al.}(2014)Madhusudhan, Amin, \&
  Kennedy}]{madhusudhan_toward_2014}
Madhusudhan, N., Amin, M.~A., \& Kennedy, G.~M. 2014, The Astrophysical Journal
  Letters, 794, L12, \dodoi{10.1088/2041-8205/794/1/L12}

\bibitem[{Mansfield {et~al.}(2021)Mansfield, Line, Bean, Fortney, Parmentier,
  Wiser, Kempton, Gharib-Nezhad, Sing, López-Morales, Baxter, Désert, Swain,
  \& Roudier}]{mansfield_unique_2021}
Mansfield, M., Line, M.~R., Bean, J.~L., {et~al.} 2021, Nature Astronomy, 5,
  1224, \dodoi{10.1038/s41550-021-01455-4}

\bibitem[{Maxted {et~al.}(2012)Maxted, Anderson, Cameron, Doyle, Fumel, Gillon,
  Hellier, Jehin, Lendl, Pepe, Pollacco, Queloz, Ségransan, Smalley,
  Southworth, Smith, Triaud, Udry, \& West}]{maxted_wasp-77_2012}
Maxted, P. F.~L., Anderson, D.~R., Cameron, A.~C., {et~al.} 2012, Publications
  of the Astronomical Society of the Pacific, 125, 48, \dodoi{10.1086/669231}

\bibitem[{Nielsen {et~al.}(2020)Nielsen, Brahm, Bouchy, Espinoza, Turner,
  Rappaport, Pearce, Ricker, Vanderspek, Latham, Seager, Winn, Jenkins, Acton,
  Bakos, Barclay, Barkaoui, Bhatti, Briceño, Bryant, Burleigh, Ciardi,
  Collins, Collins, Cooke, Csubry, Santos, Eigmüller, Fausnaugh, Gan, Gillon,
  Goad, Guerrero, Hagelberg, Hart, Henning, Huang, Jehin, Jenkins, Jordàn,
  Kielkopf, Kossakowski, Lavie, Law, Lendl, de~Leon, Lovis, Mann, Marmier,
  McCormac, Mori, Moyano, Narita, Osip, Otegi, Pepe, Pozuelos, Raynard, Relles,
  Sarkis, Segransan, Seidel, Shporer, Stalport, Stockdale, Suc, Tamura, Tan,
  Tilbrook, Ting, Trifonov, Udry, Vanderburg, Wheatley, Wingham, Zhan, \&
  Ziegler}]{nielsen_three_2020}
Nielsen, L.~D., Brahm, R., Bouchy, F., {et~al.} 2020, Astronomy \&
  Astrophysics, 639, A76, \dodoi{10.1051/0004-6361/202037941}

\bibitem[{Park {et~al.}(2014)Park, Jaffe, Yuk, Chun, Pak, Kim, Pavel, Lee, Oh,
  Jeong, Sim, Lee, Le, Strubhar, Gully-Santiago, Oh, Cha, Moon, Park, Brooks,
  Ko, Han, Nah, Hill, Lee, Barnes, Yu, Kaplan, Mace, Kim, Lee, Hwang, \&
  Park}]{park_design_2014}
Park, C., Jaffe, D.~T., Yuk, I.-S., {et~al.} 2014, in Ground-based and
  {Airborne} {Instrumentation} for {Astronomy} {V}, Vol. 9147 (SPIE), 510--521,
  \dodoi{10.1117/12.2056431}

\bibitem[{Parmentier {et~al.}(2018)Parmentier, Line, Bean, Mansfield,
  Kreidberg, Lupu, Visscher, Désert, Fortney, Deleuil, Arcangeli, Showman, \&
  Marley}]{parmentier_thermal_2018}
Parmentier, V., Line, M.~R., Bean, J.~L., {et~al.} 2018, Astronomy \&
  Astrophysics, 617, A110, \dodoi{10.1051/0004-6361/201833059}

\bibitem[{Pelletier {et~al.}(2021)Pelletier, Benneke, Darveau-Bernier, Boucher,
  Cook, Piaulet, Coulombe, Artigau, Lafrenière, Delisle, Allart, Doyon,
  Donati, Fouqué, Moutou, Cadieux, Delfosse, Hébrard, Martins, Martioli, \&
  Vandal}]{pelletier_where_2021}
Pelletier, S., Benneke, B., Darveau-Bernier, A., {et~al.} 2021, Where is the
  {Water}? {Jupiter}-like {C}/{H} ratio but strong {H}\$\_2\${O} depletion
  found on \${\textbackslash}tau\$ {Bo}{\textbackslash}"otis b using {SPIRou},
  \dodoi{10.3847/1538-3881/ac0428}

\bibitem[{Pelletier {et~al.}(2023)Pelletier, Benneke, Ali-Dib, Prinoth, Kasper,
  Seifahrt, Bean, Debras, Klein, Bazinet, Hoeijmakers, Kesseli, Lim, Carmona,
  Pino, Casasayas-Barris, Hood, \& Stürmer}]{pelletier_vanadium_2023}
Pelletier, S., Benneke, B., Ali-Dib, M., {et~al.} 2023, Nature, 1,
  \dodoi{10.1038/s41586-023-06134-0}

\bibitem[{Polyansky {et~al.}(2018)Polyansky, Kyuberis, Zobov, Tennyson,
  Yurchenko, \& Lodi}]{polyansky_exomol_2018}
Polyansky, O.~L., Kyuberis, A.~A., Zobov, N.~F., {et~al.} 2018, Monthly Notices
  of the Royal Astronomical Society, 480, 2597, \dodoi{10.1093/mnras/sty1877}

\bibitem[{Reiners \& Schmitt(2002)}]{reiners_feasibility_2002}
Reiners, A., \& Schmitt, J. H. M.~M. 2002, Astronomy \& Astrophysics, 384, 155,
  \dodoi{10.1051/0004-6361:20011801}

\bibitem[{Robitaille {et~al.}(2013)Robitaille, Tollerud, Greenfield,
  Droettboom, Bray, Aldcroft, Davis, Ginsburg, Price-Whelan, Kerzendorf,
  Conley, Crighton, Barbary, Muna, Ferguson, Grollier, Parikh, Nair, Günther,
  Deil, Woillez, Conseil, Kramer, Turner, Singer, Fox, Weaver, Zabalza,
  Edwards, Bostroem, Burke, Casey, Crawford, Dencheva, Ely, Jenness, Labrie,
  Lim, Pierfederici, Pontzen, Ptak, Refsdal, Servillat, \&
  Streicher}]{robitaille_astropy_2013}
Robitaille, T.~P., Tollerud, E.~J., Greenfield, P., {et~al.} 2013, Astronomy \&
  Astrophysics, 558, A33, \dodoi{10.1051/0004-6361/201322068}

\bibitem[{Rothman {et~al.}(2010)Rothman, Gordon, Barber, Dothe, Gamache,
  Goldman, Perevalov, Tashkun, \& Tennyson}]{rothman_hitemp_2010}
Rothman, L.~S., Gordon, I.~E., Barber, R.~J., {et~al.} 2010, Journal of
  Quantitative Spectroscopy and Radiative Transfer, 111, 2139,
  \dodoi{10.1016/j.jqsrt.2010.05.001}

\bibitem[{Smith {et~al.}(2024)Smith, Line, Bean, Brogi, August, Welbanks,
  Desert, Lunine, Sanchez, Mansfield, Pino, Rauscher, Kempton, Zalesky, \&
  Fowler}]{smith_combined_2024}
Smith, P. C.~B., Line, M.~R., Bean, J.~L., {et~al.} 2024, The Astronomical
  Journal, 167, 110, \dodoi{10.3847/1538-3881/ad17bf}

\bibitem[{Snellen {et~al.}(2010)Snellen, de~Kok, de~Mooij, \&
  Albrecht}]{snellen_orbital_2010}
Snellen, I. A.~G., de~Kok, R.~J., de~Mooij, E. J.~W., \& Albrecht, S. 2010,
  Nature, 465, 1049, \dodoi{10.1038/nature09111}

\bibitem[{Stock {et~al.}(2022)Stock, Kitzmann, \& Patzer}]{stock_fastchem_2022}
Stock, J.~W., Kitzmann, D., \& Patzer, A. B.~C. 2022, {FastChem} 2: {An}
  improved computer program to determine the gas-phase chemical equilibrium
  composition for arbitrary element distributions,
  \dodoi{10.1093/mnras/stac2623}

\bibitem[{Stock {et~al.}(2018)Stock, Kitzmann, Patzer, \&
  Sedlmayr}]{stock_fastchem_2018}
Stock, J.~W., Kitzmann, D., Patzer, A. B.~C., \& Sedlmayr, E. 2018, Monthly
  Notices of the Royal Astronomical Society, 479, 865,
  \dodoi{10.1093/mnras/sty1531}

\bibitem[{Thorngren \& Fortney(2019)}]{thorngren_connecting_2019}
Thorngren, D., \& Fortney, J.~J. 2019, The Astrophysical Journal Letters, 874,
  L31, \dodoi{10.3847/2041-8213/ab1137}

\bibitem[{Virtanen {et~al.}(2020)Virtanen, Gommers, Oliphant, Haberland, Reddy,
  Cournapeau, Burovski, Peterson, Weckesser, Bright, van~der Walt, Brett,
  Wilson, Millman, Mayorov, Nelson, Jones, Kern, Larson, Carey, Polat, Feng,
  Moore, VanderPlas, Laxalde, Perktold, Cimrman, Henriksen, Quintero, Harris,
  Archibald, Ribeiro, Pedregosa, \& van Mulbregt}]{virtanen_scipy_2020}
Virtanen, P., Gommers, R., Oliphant, T.~E., {et~al.} 2020, Nature Methods, 17,
  261, \dodoi{10.1038/s41592-019-0686-2}

\bibitem[{Wong {et~al.}(2020)Wong, Shporer, Daylan, Benneke, Fetherolf, Kane,
  Ricker, Vanderspek, Latham, Winn, Jenkins, Boyd, Glidden, Goeke, Sha, Ting,
  \& Yahalomi}]{wong_systematic_2020}
Wong, I., Shporer, A., Daylan, T., {et~al.} 2020, The Astronomical Journal,
  160, 155, \dodoi{10.3847/1538-3881/ababad}

\bibitem[{Yurchenko {et~al.}(2020)Yurchenko, Mellor, Freedman, \&
  Tennyson}]{yurchenko_exomol_2020}
Yurchenko, S.~N., Mellor, T.~M., Freedman, R.~S., \& Tennyson, J. 2020, Monthly
  Notices of the Royal Astronomical Society, 496, 5282,
  \dodoi{10.1093/mnras/staa1874}

\bibitem[{Yurchenko {et~al.}(2022)Yurchenko, Tennyson, Syme, Adam, Clark,
  Cooper, Dobney, Donnelly, Gorman, Lynas-Gray, Meltzer, Owens, Qu, Semenov,
  Somogyi, Upadhyay, Wright, \& Zapata Trujillo}]{yurchenko_exomol_2022}
Yurchenko, S.~N., Tennyson, J., Syme, A.-M., {et~al.} 2022, Monthly Notices of
  the Royal Astronomical Society, 510, 903, \dodoi{10.1093/mnras/stab3267}

\bibitem[{Öberg \& Bergin(2016)}]{oberg_excess_2016}
Öberg, K.~I., \& Bergin, E.~A. 2016, The Astrophysical Journal Letters, 831,
  L19, \dodoi{10.3847/2041-8205/831/2/L19}

\bibitem[{Öberg {et~al.}(2011)Öberg, Murray-Clay, \&
  Bergin}]{oberg_effects_2011}
Öberg, K.~I., Murray-Clay, R., \& Bergin, E.~A. 2011, The Astrophysical
  Journal Letters, 743, L16, \dodoi{10.1088/2041-8205/743/1/L16}

\end{thebibliography}
\bibliographystyle{aasjournal}



\end{document}